\documentclass{article}

\usepackage{arxiv}

\usepackage{threeparttable}
\usepackage[utf8]{inputenc} 
\usepackage[T1]{fontenc}    
\usepackage{hyperref}       
\usepackage{url}   
\usepackage{natbib}
\usepackage{amsmath}
\usepackage{booktabs}       
\usepackage{amsfonts}       
\usepackage{nicefrac}       
\usepackage{microtype}      
\usepackage{lipsum}
\usepackage{graphicx}
\usepackage{nicefrac}       
\usepackage{algorithm}
\usepackage{algpseudocode}
\graphicspath{ {./images/} }

\title{Learning Interpretable Text Signals for Structured Responses}

\author{
Cixiao Jiang\thanks{Corresponding author. ORCID: 0000-0002-1566-9191. Email: \href{mailto:cixiao.jiang@york.ac.uk}{cixiao.jiang@york.ac.uk}.} \\
Department of Mathematics\\
University of York\\
York, UK
\and
Ben Powell\\
Department of Mathematics\\
University of York\\
York, UK
\and
Niall MacKay\\
Department of Mathematics\\
University of York\\
York, UK
}

\date{}

\begin{document}
\maketitle
\begin{abstract}
Textual data are often collected alongside structured response variables, but prediction and interpretation are commonly treated as separate tasks. This paper studies rating prediction as an initial case of interpretable text-response modelling, where the aim is to learn textual representations that are both semantically meaningful and aligned with an external response. We propose a joint non-negative matrix factorisation and binomial regression model, in which the document-topic representation is learned from both text reconstruction and rating prediction. Simulation experiments and a real-world review dataset show that the model can recover stable response-relevant textual signals and achieve competitive performance against linear and ridge regression baselines. The framework provides a practical step towards interpretable modelling of text-linked outcomes, with potential extensions to other response types beyond bounded ratings.

\end{abstract}

\section{Introduction}
\label{sec:introduction}

Textual data are often collected alongside structured response variables. In biomedical research, clinical notes, adverse-event narratives and patient-reported comments may be linked to severity grades, symptom scores, treatment responses or risk levels. In survey and policy research, open-ended responses are frequently paired with Likert-scale attitudes or ordered assessments. Similar structures appear in education, finance, geopolitical risk analysis and online platforms, where textual explanations accompany numerical ratings, risk categories or performance indicators. Online reviews are one familiar example of this broader data structure: they combine free-text descriptions of experience with bounded numerical ratings and both review content and ratings have been linked to demand, sales and market outcomes \citep{chevalier2006effect,luca2016reviews,archak2011deriving}.

Across these settings, the analytical goal is rarely prediction alone. Researchers and domain experts often need to understand which textual patterns are associated with variation in the response. A low rating, high risk score or severe clinical grade has different implications depending on whether it is linked to service quality, safety concerns, treatment symptoms, financial uncertainty or other latent dimensions in the text. The value of text-response modelling therefore lies in learning representations that are both predictive and interpretable: the model should connect textual evidence with the external response while preserving a meaningful semantic structure that can support substantive explanation.

The statistical difficulty is that the two sides of this data structure have very different forms. The response variable is structured and directly usable, but it may be bounded, discrete, skewed or otherwise constrained. Online ratings, for example, are often concentrated near the upper end of the scale \citep{hu2009overcoming}. By contrast, the text contains the substantive evidence needed to explain the response, but it is sparse, high-dimensional and heterogeneous \citep{pang2008opinion,yan2013biterm}. This mismatch becomes more pronounced in small-sample settings, where the number of documents may be limited while the vocabulary remains large. Although the broader motivation is to link text with structured responses of different types, this paper focuses on bounded ratings as an initial and practically important case.

Several existing approaches address parts of this problem. Sentiment analysis provides a transparent and widely used way to summarise subjective language, and lexicon-based tools remain useful for exploratory text analysis \citep{pang2008opinion,liu2012sentiment,hutto2014vader,pennebaker2015liwc}. However, a single sentiment score is often too coarse to explain variation in a structured response. A review, clinical note or survey comment may contain several distinct dimensions at once, and reducing it to an overall positive or negative score removes much of the structure that the analyst wants to inspect. More flexible methods, including transformer-based language models, can learn richer contextual representations \citep{vaswani2017attention,devlin2019bert}, but their predictive features are not always easy to translate into stable, concrete explanations \citep{lipton2018mythos,jacovi2020towards}.

A direct predictive alternative is to regress the response on word-level features such as term frequency-inverse document frequency (TF-IDF) weights \citep{salton1988term}. Regularised linear models, including Ridge and Lasso regression, provide strong baselines because they are simple, stable and effective in high-dimensional settings \citep{hoerl1970ridge,tibshirani1996regression,hastie2009elements}. Their limitation is not that they are weak predictors, but that their interpretation remains largely word-based. A coefficient on an individual term may indicate association with the response, but it does not by itself define a broader semantic dimension. In many applications, decisions and explanations are required at the level of latent aspects, themes or mechanisms rather than isolated words.

Topic models provide this intermediate layer of interpretation. Classical models such as latent Dirichlet allocation (LDA) and non-negative matrix factorisation (NMF) represent documents through low-dimensional latent topics or components, allowing related terms to be grouped into interpretable semantic dimensions \citep{blei2003latent,griffiths2004finding,lee1999learning}. Topic-based approaches have been widely used for review analysis and aspect discovery \citep{titov2008modeling,lu2011multi}. However, standard unsupervised topic models are fitted to explain textual variation, not variation in an external response. 
A coherent topic is therefore not necessarily a response-relevant topic. Related work on supervised topic models, structural topic models and inverse regression for text shows that topic structure can be linked to document-level responses or covariates \citep{mcauliffe2008supervised,roberts2014structural,taddy2013multinomial}. 
This motivates a response-aware formulation in which the textual representation is learned jointly with the response model rather than estimated in a separate first stage.

This paper starts from this gap. The aim is not simply to discover coherent topics, nor simply to maximise predictive accuracy from high-dimensional word features. Instead, the aim is to learn a low-dimensional textual representation that is both semantically interpretable and statistically aligned with an external response. In the present paper, the response is a bounded rating, but the broader modelling idea is text-response learning: textual patterns should be connected to a structured response variable through a shared representation. This requires balancing three objectives: preserving the semantic structure of the text, extracting response-relevant variation, and controlling model complexity under sparse or small-sample conditions.

We propose a Joint Binomial Nonnegative Matrix Factorisation (JBNMF) model for this purpose. The model links a non-negative topic representation of the TF-IDF document-term matrix with a binomial working likelihood for bounded five-star ratings, so that text reconstruction and rating prediction are estimated under a single objective. The contribution is not a claim of universal predictive superiority but a supervised matrix-factorisation framework that offers competitive prediction with an inspectable topic layer. The paper makes three main contributions: it formulates rating prediction as an initial case of interpretable text-response modelling; it develops a joint NMF and binomial regression objective for learning response-aware textual representations; and it studies the trade-off between reconstruction, prediction and regularisation through simulation and real-world review data. Section~\ref{methodology} defines the model and estimation procedure; Section~\ref{sec:simulation} studies its behaviour in controlled simulations; Section~\ref{sec:application} evaluates it on online review data; and Section~\ref{sec:discussion} summarises the main implications and limitations.

\section{Methodology}\label{methodology}

\subsection{Text representation}
\label{sec:text_representation}

The proposed model starts by converting unstructured review text into a structured numerical input. This step is needed because the subsequent matrix-factorisation model operates on a document-term matrix rather than on raw text. We use TF-IDF rather than a dense semantic embedding because the aim is to retain a sparse, nonnegative and term-level representation that is directly compatible with nonnegative matrix factorisation and remains easy to inspect \citep{salton1988term}.

Let \(i=1,\ldots,M\) index reviews and \(j=1,\ldots,N\) index retained terms. Here \(M\) is the number of reviews and \(N\) is the effective vocabulary size after preprocessing, TF-IDF vectorisation and vocabulary filtering. The filtering step removes terms that are either too rare to provide stable information or too common to be informative. For review \(i\) and retained term \(j\), let \(c_{ij}\) denote the raw count of term \(j\) in review \(i\). Since each review contains only a small subset of the retained vocabulary, the count matrix is typically sparse.

To reduce the influence of common but weakly informative words, we apply TF-IDF weighting. Following the smoothed weighting convention used by the \texttt{scikit-learn} package in Python, the term frequency is defined as
\[
\mathrm{tf}_{ij}
=
\frac{c_{ij}}{\sum_{v=1}^{N} c_{iv}},
\]
and the inverse document frequency is defined as
\[
\mathrm{idf}_{j}
=
\log\left\{\frac{1+M}{1+m_j}\right\}+1,
\]
where \(m_j\) is the number of reviews in which retained term \(j\) appears. 
The additive constants follow the smoothed TF-IDF convention used in our implementation: the smoothing inside the logarithm stabilises the document-frequency ratio, while the final \(+1\) ensures that terms appearing in all reviews retain a positive inverse-document-frequency weight. The final document-term entry is
\begin{equation}
X_{ij}
=
\mathrm{tf}_{ij}\mathrm{idf}_{j}.
\label{eq:tfidf}
\end{equation}

This gives a nonnegative matrix \(\mathbf{X}\in\mathbb{R}^{M\times N}_{+}\), where row \(i\) represents review \(i\) and column \(j\) represents retained term \(j\). A larger value of \(X_{ij}\) indicates that term \(j\) is frequent in review \(i\) but relatively uncommon across the corpus. The model treats \(\mathbf{X}\) as the observed textual input and learns a lower-dimensional topic representation from this sparse nonnegative matrix.

\subsection{Model formulation}
\label{sec:model_formulation}

The proposed model combines two components: a nonnegative matrix factorisation model for textual representation and a binomial regression model for rating prediction. The central idea is that the same low-dimensional representation should both reconstruct the review text and explain variation in the observed rating.

\subsubsection{Nonnegative matrix factorisation}
\label{sec:nmf_formulation}

For the textual component of the model, we use nonnegative matrix factorisation (NMF) as the basic representation-learning framework. This choice is motivated by the structure of the review data considered in this paper. After TF-IDF vectorisation, the observed text is represented by a nonnegative, high-dimensional and sparse document-term matrix. In addition, the number of available reviews may be limited and each individual review is usually short. Under these conditions, it is useful to adopt a text model that can learn a low-dimensional semantic representation directly from the observed document-term matrix.

NMF is suitable for this purpose because it represents the text matrix through additive nonnegative components. Unlike latent Dirichlet allocation (LDA), which specifies a probabilistic generative model and infers document-level topic mixtures from word assignments \citep{blei2003latent}, NMF works directly with the nonnegative matrix representation. This avoids the need to estimate a full document-level topic distribution from very short documents, where each review may contain only limited word co-occurrence information. Previous work has also used NMF for document clustering and text representation, showing that the learned nonnegative components can be interpreted as base textual patterns or topics \citep{lee2001algorithms,xu2003document}. In this paper, NMF is therefore used as a stable and interpretable foundation for learning the latent textual structure that will subsequently be linked to the rating outcome.

Let $\mathbf{X}\in\mathbb{R}^{M\times N}_{+}$ denote the TF-IDF document-term matrix defined in Section~\ref{sec:text_representation}. NMF approximates this matrix by a low-rank nonnegative factorisation,
\begin{equation}
\mathbf{X}
\approx
\mathbf{W}\mathbf{H},
\label{eq:nmf_factorisation}
\end{equation}
where \(\mathbf{W}\in\mathbb{R}^{M\times K}_{+}\) is the document–topic matrix, \(\mathbf{H}\in\mathbb{R}^{K\times N}_{+}\) is the topic–term matrix and \(K\) is the number of latent topics. The approximation sign indicates that the observed document-term matrix is reconstructed by a lower-rank nonnegative representation, rather than matched exactly. The row vector \(\mathbf{W}_{i\cdot}\) represents the topic mixture of review \(i\), while \(\mathbf{H}_{k\cdot}\) represents the term profile of topic \(k\). Topic interpretation is based on the rows of \(\mathbf{H}\) and the rating model below uses the rows of \(\mathbf{W}\) as document-level predictors.

The textual part of the model is learned using the generalised Kullback-Leibler divergence commonly used in KL-NMF \citep{lee2001algorithms}. This objective is not used here as a probabilistic KL divergence between two normalised distributions but as the standard NMF reconstruction loss for nonnegative document-term matrices. The text-learning loss is defined as
\begin{equation}
\mathcal{L}_{\mathrm{text}} = D_{\mathrm{KL}}(\mathbf{X}\Vert \mathbf{W}\mathbf{H})=\sum_{i=1}^{M}\sum_{j=1}^{N}\left[X_{ij}\log\frac{X_{ij}}{(\mathbf{W}\mathbf{H})_{ij}}-X_{ij}
+(\mathbf{W}\mathbf{H})_{ij}\right],
\label{eq:text_loss}
\end{equation}
where entries with \(X_{ij}=0\) are evaluated using the standard limiting convention \(X_{ij}\log\{X_{ij}/(\mathbf{W}\mathbf{H})_{ij}\}=0\). In numerical implementation, a small positive constant is added to \((\mathbf{W}\mathbf{H})_{ij}\) to avoid division by zero. This loss encourages \(\mathbf{W}\) and \(\mathbf{H}\) to preserve the main word-use structure in the review corpus while maintaining a nonnegative and interpretable low-dimensional representation. In the joint model developed below, \(\mathcal{L}_{\mathrm{text}}\) is combined with a rating-learning loss so that the learned document representation is informed by both textual structure and rating information.

\subsubsection{Binomial rating model}
\label{sec:binomial_rating_model}

Star ratings are ordered, discrete and bounded responses. Treating them with a standard Gaussian regression model would ignore both the finite support of the outcome and the ordinal structure of the scale. Classical ordinal regression models address this type of response by modelling cumulative response probabilities, thereby using the ordering of the categories without assuming that the numerical labels are fully cardinal measurements \citep{mccullagh1980regression}.

Here we use a more parsimonious shifted-binomial working likelihood. This is not intended as a literal claim that users generate ratings through four independent Bernoulli trials. Rather, the shifted-binomial form provides a one-probability representation for a five-level bounded rating, which can be coupled directly to the document-topic representation through \(\mathbf{W}_{i\cdot}\beta\).

Let $r_i\in\{1,2,3,4,5\}$ denote the observed star rating for review $i$. We transform the rating into a bounded count by setting
\begin{equation}
Y_i
=
r_i-1,
\qquad
Y_i\in\{0,1,2,3,4\}.
\label{eq:rating_transform}
\end{equation}
The transformed rating is modelled as
\begin{equation}
Y_i
\sim
\mathrm{Binomial}(4,p_i),
\label{eq:binomial_rating}
\end{equation}
where $p_i$ is the success probability associated with review $i$. Under this representation, the lowest rating corresponds to zero successes and the highest rating corresponds to four successes.

The success probability is linked to the latent textual representation through a logistic regression model:
\begin{equation}
p_i
=
\sigma(\mathbf{W}_{i\cdot}\boldsymbol{\beta}),
\qquad
\sigma(z)
=
\frac{1}{1+\exp(-z)},
\label{eq:binomial_link}
\end{equation}
where $\boldsymbol{\beta}\in\mathbb{R}^{K}$ is the regression coefficient vector. This follows the binomial generalised linear modelling framework with a logit link \citep{mccullagh1989generalized,agresti2015foundations}. The difference from a conventional text regression model is that the covariates are not the original high-dimensional TF-IDF terms but the latent document-topic weights learned through NMF. The coefficient $\beta_k$ therefore measures the association between topic $k$ and the rating tendency on the logit scale.

The rating-learning loss is defined as the negative binomial log-likelihood. The full binomial likelihood contains the coefficient $\binom{4}{Y_i}$ but this term does not depend on the model parameters and can be omitted from the optimisation objective. The resulting rating-learning loss is
\begin{equation}
\mathcal{L}_{\mathrm{rating}}
=
-
\sum_{i=1}^{M}
\left[
Y_i\log p_i
+
(4-Y_i)\log(1-p_i)
\right].
\label{eq:rating_loss}
\end{equation}
This term measures how well the latent document representation $\mathbf{W}$ and the regression coefficients $\boldsymbol{\beta}$ explain the observed star ratings. In the joint model, $\mathcal{L}_{\mathrm{rating}}$ is combined with the text-learning loss so that the latent representation is informed not only by word-use patterns but also by rating-relevant variation in the reviews.

\subsubsection{Joint Binomial NMF}
\label{sec:joint_binomial_nmf}

The Joint Binomial NMF model combines the text-learning loss in Eq.~\eqref{eq:text_loss} with the rating-learning loss in Eq.~\eqref{eq:rating_loss}. The model is estimated by minimising
\begin{equation}
\mathcal{L}_{\mathrm{JBNMF}}
=
\mathcal{L}_{\mathrm{text}}
+
\alpha
\left[
\mathcal{L}_{\mathrm{rating}}
+
\lambda \Vert \boldsymbol{\beta} \Vert_2^2
\right],
\label{eq:jbnmf_objective}
\end{equation}
subject to
\begin{equation}
\mathbf{W}\geq 0,
\qquad
\mathbf{H}\geq 0.
\label{eq:nonnegative_constraints}
\end{equation}

\begin{figure}[!t]%
\centering
\includegraphics[width=0.82\linewidth]{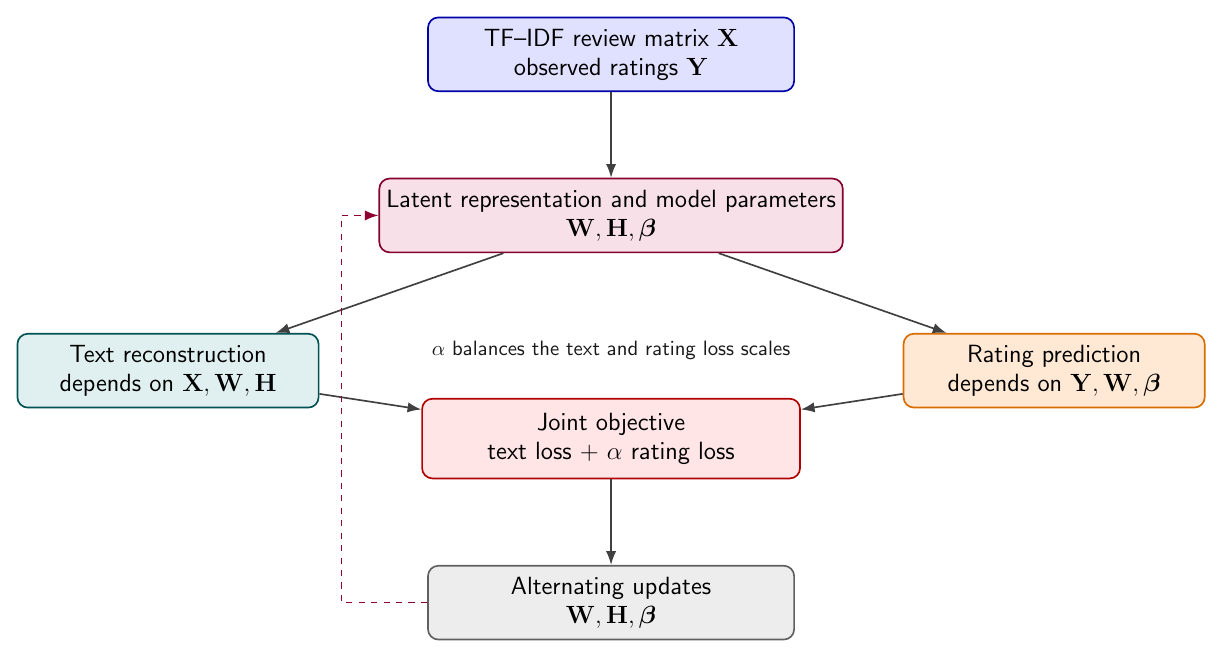}
\caption{
Schematic structure of the JBNMF model. The document representation \(\mathbf{W}\) links the KL-NMF text-reconstruction component with the binomial rating-prediction component, while \(\alpha\) balances the numerical contribution of the two losses.
}
\label{fig:jbnmf_process}
\end{figure}

The objective in Eq.~\eqref{eq:jbnmf_objective} has three parts. The first term, 
$\mathcal{L}_{\mathrm{text}}$, learns the latent semantic structure of the review corpus through KL-NMF. The second term, $\mathcal{L}_{\mathrm{rating}}$, links the document-level topic weights to the observed star ratings through the binomial rating model. The third term, $\lambda\Vert\boldsymbol{\beta}\Vert_2^2$, regularises the rating component by penalising large regression coefficients.

The parameter $\alpha\geq 0$ is a loss-balancing parameter that controls the relative numerical contribution of the rating likelihood to the joint objective. When $\alpha=0$, the model reduces to unsupervised KL-NMF and the rating information does not affect the learned representation. When $\alpha>0$, the rating-learning loss influences the document-topic matrix $\mathbf{W}$, encouraging it to capture textual variation that is predictive of ratings as well as variation that explains the document-term matrix. The parameter $\lambda\geq 0$ controls the strength of the $L_2$ penalty on $\boldsymbol{\beta}$.

The central feature of the formulation is therefore the shared role of $\mathbf{W}$. Rather than first extracting topics and then fitting a separate prediction model, JBNMF estimates a latent representation that is shaped jointly by word-use patterns and rating-relevant variation. This makes the model a supervised representation-learning framework for sparse review text, rather than a two-stage topic-modelling and regression procedure. The overall optimisation structure is summarised in Figure~\ref{fig:jbnmf_process}.
\subsection{Parameter estimation}
\label{sec:parameter_estimation}

The objective in Eq.~\eqref{eq:jbnmf_objective} is nonconvex in $(\mathbf{W},\mathbf{H},\boldsymbol{\beta})$ because of the bilinear factorisation $\mathbf{W}\mathbf{H}$ and the regression term $\mathbf{W}\boldsymbol{\beta}$. As in standard NMF problems, this motivates an alternating block-coordinate estimation strategy \citep{lee2001algorithms,kim2014algorithms}. We therefore estimate the model using an alternating block-coordinate procedure. At each iteration, $\mathbf{H}$, $\boldsymbol{\beta}$ and $\mathbf{W}$ are updated in turn while holding the remaining parameter blocks fixed.

Given $\mathbf{W}$, the topic-term matrix $\mathbf{H}$ is updated using the standard multiplicative update for KL-NMF:
\begin{equation}
\mathbf{H}
\leftarrow
\mathbf{H}\odot
\frac{
\mathbf{W}^{\top}
\left[
\mathbf{X}/(\mathbf{W}\mathbf{H})
\right]
}{
\mathbf{W}^{\top}\mathbf{1}_{M}
},
\label{eq:update_h}
\end{equation}
where $\odot$ denotes element-wise multiplication, the division is element-wise and $\mathbf{1}_{M}$ is a column vector of ones of length $M$.

Given $\mathbf{W}$, the regression coefficients $\boldsymbol{\beta}$ are updated by a gradient step for the regularised binomial likelihood. Let $\mathbf{p}=(p_1,\ldots,p_M)^{\top}$ and $\mathbf{Y}=(Y_1,\ldots,Y_M)^{\top}$. The update is
\begin{equation}
\boldsymbol{\beta}
\leftarrow
\boldsymbol{\beta}
-
\eta\alpha
\left[
\mathbf{W}^{\top}(4\mathbf{p}-\mathbf{Y})
+
2\lambda\boldsymbol{\beta}
\right],
\label{eq:update_beta}
\end{equation}
where $\eta>0$ is the learning rate.

The document-topic matrix $\mathbf{W}$ is shared by the text-learning loss and the rating-learning loss. It therefore plays a dual role: it must provide a nonnegative low-dimensional approximation to the document-term matrix, while also carrying rating-relevant information through the binomial regression model. To update $\mathbf{W}$ under these two requirements, we use a supervised multiplicative update that combines the KL-NMF contribution from $\mathcal{L}_{\mathrm{text}}$ with the gradient contribution from $\mathcal{L}_{\mathrm{rating}}$. Define
\[
G_{ik}
=
(4p_i-Y_i)\beta_k .
\]
Then
\begin{equation}
W_{ik}
\leftarrow
W_{ik}
\frac{
\sum_{j=1}^{N}
H_{kj}
\frac{X_{ij}}{(\mathbf{W}\mathbf{H})_{ij}}
+
\alpha \max(0,-G_{ik})
+
\gamma
}{
\sum_{j=1}^{N} H_{kj}
+
\alpha \max(0,G_{ik})
+
\gamma
},
\label{eq:update_w}
\end{equation}
where $\gamma>0$ is a stabilising constant. This update preserves the nonnegativity constraint on $\mathbf{W}$ while allowing the binomial rating component to influence the learned document-topic representation.

The three updates are repeated until the relative change in the objective function falls below a predefined tolerance or until a maximum number of iterations is reached. The detailed derivation of Eqs.~\eqref{eq:update_h}--\eqref{eq:update_w} is provided in Appendix~\ref{app:update_derivations}.

The estimation procedure is summarised in Algorithm~\ref{alg:jbnmf_training}. 
In this algorithm, $\odot$, division and the maximum operator are applied element-wise.

\begin{algorithm}[!htbp]
\small
\caption{Joint Binomial NMF model}
\label{alg:jbnmf_training}
\begin{algorithmic}[1]

\Require Document-term matrix $\mathbf{X}$, shifted ratings $\mathbf{Y}$, number of topics $K$, tuning parameters $\alpha$ and $\lambda$
\Ensure Fitted parameters $\widehat{\mathbf{W}}$, $\widehat{\mathbf{H}}$ and $\widehat{\boldsymbol{\beta}}$

\State Initialise $\mathbf{W}\in\mathbb{R}^{M\times K}_{+}$, $\mathbf{H}\in\mathbb{R}^{K\times N}_{+}$ and $\boldsymbol{\beta}\in\mathbb{R}^{K}$
\State Choose $\eta>0$, $\gamma>0$, $\varepsilon>0$, tolerance $\tau$ and maximum iterations $T_{\max}$
\State Compute $\mathcal{L}^{(0)}_{\mathrm{JBNMF}}$ using Eq.~\eqref{eq:jbnmf_objective}

\For{$t=1,\ldots,T_{\max}$}

    \State $\mathbf{R}\leftarrow \mathbf{X}/(\mathbf{W}\mathbf{H}+\varepsilon)$

    \State $\mathbf{H}\leftarrow
    \mathbf{H}\odot
    \dfrac{\mathbf{W}^{\top}\mathbf{R}}
    {\mathbf{W}^{\top}\mathbf{1}_{M}+\varepsilon}$

    \State $\mathbf{p}\leftarrow \sigma(\mathbf{W}\boldsymbol{\beta})$

    \State $\boldsymbol{\beta}\leftarrow
    \boldsymbol{\beta}
    -
    \eta\alpha
    \left[
    \mathbf{W}^{\top}(4\mathbf{p}-\mathbf{Y})
    +
    2\lambda\boldsymbol{\beta}
    \right]$

    \State $\mathbf{p}\leftarrow \sigma(\mathbf{W}\boldsymbol{\beta})$
    \State $\mathbf{R}\leftarrow \mathbf{X}/(\mathbf{W}\mathbf{H}+\varepsilon)$

    \State $\mathbf{q}\leftarrow 4\mathbf{p}-\mathbf{Y}$, \quad
    $\mathbf{G}\leftarrow \mathbf{q}\boldsymbol{\beta}^{\top}$

    \State $\mathbf{G}^{+}\leftarrow \max(\mathbf{G},0)$, \quad
    $\mathbf{G}^{-}\leftarrow \max(-\mathbf{G},0)$

    \State $\mathbf{A}\leftarrow
    \mathbf{R}\mathbf{H}^{\top}
    +
    \alpha\mathbf{G}^{-}
    +
    \gamma\mathbf{1}_{M\times K}$

    \State $\mathbf{B}\leftarrow
    \mathbf{1}_{M}(\mathbf{H}\mathbf{1}_{N})^{\top}
    +
    \alpha\mathbf{G}^{+}
    +
    \gamma\mathbf{1}_{M\times K}$

    \State $\mathbf{W}\leftarrow \mathbf{W}\odot \mathbf{A}/\mathbf{B}$

    \State Compute $\mathcal{L}^{(t)}_{\mathrm{JBNMF}}$ using Eq.~\eqref{eq:jbnmf_objective}

    \If{$
    \left|
    \mathcal{L}^{(t)}_{\mathrm{JBNMF}}
    -
    \mathcal{L}^{(t-1)}_{\mathrm{JBNMF}}
    \right|
    /
    \max\{1,\left|\mathcal{L}^{(t-1)}_{\mathrm{JBNMF}}\right|\}
    < \tau
    $}
        \State \textbf{break}
    \EndIf

\EndFor

\State $\widehat{\mathbf{W}}\leftarrow \mathbf{W}$, 
$\widehat{\mathbf{H}}\leftarrow \mathbf{H}$ and 
$\widehat{\boldsymbol{\beta}}\leftarrow \boldsymbol{\beta}$

\end{algorithmic}
\end{algorithm}

\subsection{Out-of-sample prediction}
\label{sec:out_of_sample_prediction}

The fitted model is used to predict the rating of a new review from its text. The prediction target is the shifted rating $Y_i=r_i-1$, with $Y_i\in\{0,\ldots,4\}$, or equivalently the original star rating $r_i\in\{1,\ldots,5\}$. After training, the model provides the fitted topic-term matrix $\widehat{\mathbf{H}}$ and the fitted regression coefficients $\widehat{\boldsymbol{\beta}}$. For a new review, however, the document-topic representation is unknown. The key out-of-sample step is therefore to infer $\widehat{\mathbf{W}}_{\mathrm{test}}$ from the new review text, using the fixed topic-term matrix $\widehat{\mathbf{H}}$.

The test reviews are transformed using the vocabulary and TF-IDF weights fitted on the training set. Therefore, the columns of $\mathbf{X}_{\mathrm{test}}$ correspond exactly to the retained training vocabulary, rather than to a vocabulary re-estimated from the test set. The test reviews are transformed using the training vocabulary and IDF weights, giving $\mathbf{X}_{\mathrm{test}}\in\mathbb{R}^{M_{\mathrm{test}}\times N}_{+}$ aligned with $\widehat{\mathbf{H}}\in\mathbb{R}^{K\times N}_{+}$; terms absent from the training vocabulary are ignored. Since $\widehat{\mathbf{H}}$ is generally rectangular and the NMF approximation is not an exact linear system, $\mathbf{W}_{\mathrm{test}}$ is not obtained by matrix inversion. Instead, it is inferred by solving a fixed-$\widehat{\mathbf{H}}$ nonnegative KL-NMF problem:
\begin{equation}
\widehat{\mathbf{W}}_{\mathrm{test}}
=
\arg\min_{\mathbf{W}\geq 0}
D_{\mathrm{KL}}
\left(
\mathbf{X}_{\mathrm{test}}
\Vert
\mathbf{W}\widehat{\mathbf{H}}
\right).
\label{eq:test_w_inference}
\end{equation}
This step uses only the test review text and does not use the test ratings. In practice, Eq.~\eqref{eq:test_w_inference} is solved using the multiplicative update
\begin{equation}
\mathbf{W}_{\mathrm{test}}
\leftarrow
\mathbf{W}_{\mathrm{test}}
\odot
\frac{
\left[
\mathbf{X}_{\mathrm{test}}/
(\mathbf{W}_{\mathrm{test}}\widehat{\mathbf{H}})
\right]
\widehat{\mathbf{H}}^{\top}
}{
\mathbf{1}_{M_{\mathrm{test}}}\widehat{\mathbf{H}}^{\top}
},
\label{eq:test_w_update}
\end{equation}
where $M_{\mathrm{test}}$ is the number of test reviews and $\mathbf{1}_{M_{\mathrm{test}}}$ is a column vector of ones of length $M_{\mathrm{test}}$. The derivation follows the fixed-factor KL-NMF update described in Appendix~\ref{app:update_derivations}.

Once $\widehat{\mathbf{W}}_{\mathrm{test}}$ has been inferred, the fitted binomial rating model maps each test review to a predicted success probability:
\begin{equation}
\widehat{p}_i
=
\sigma
\left(
\widehat{\mathbf{W}}_{i\cdot,\mathrm{test}}
\widehat{\boldsymbol{\beta}}
\right),
\label{eq:test_probability}
\end{equation}
which is the out-of-sample version of Eq.~\eqref{eq:binomial_link}. The predictive distribution of the shifted rating is then
\begin{equation}
\Pr(\widehat{Y}_i=y)
=
\binom{4}{y}
\widehat{p}_i^{y}
(1-\widehat{p}_i)^{4-y},
\qquad
y\in\{0,\ldots,4\}.
\label{eq:predictive_binomial}
\end{equation}
Equivalently, the model assigns a categorical probability to each of the five shifted rating levels, with the category probabilities constrained by the binomial form.

For regression evaluation, we use the conditional expected star rating implied by the predictive distribution:
\begin{equation}
\widehat{r}_i
=
1+\mathbb{E}(\widehat{Y}_i)
=
1+4\widehat{p}_i.
\label{eq:expected_rating_prediction}
\end{equation}
For discrete rating prediction, we use the fitted binomial mean \(4\widehat{p}_i\) as the continuous prediction of the shifted rating \(Y_i=r_i-1\). This value is then mapped to the nearest admissible integer in \(\{0,1,2,3,4\}\), giving
\begin{equation}
\widehat{r}_i
=
1+\left\lfloor 4\widehat{p}_i+\frac{1}{2}\right\rfloor .
\label{eq:discrete_rating_prediction}
\end{equation}
In implementation, the resulting predictions are clipped to the original rating range \(\{1,\ldots,5\}\) to avoid numerical boundary issues.
Alternatively, a maximum-probability discrete prediction can be obtained by selecting the mode of the predictive binomial distribution in Eq.~\eqref{eq:predictive_binomial}.

\section{Simulation study}\label{sec:simulation}

\subsection{Data generation}

The simulation study is not intended to reproduce all features of online reviews. Instead, it provides a controlled setting in which the rating signal is known to arise from latent topic proportions. This setting therefore directly examines the main modelling assumption of the proposed method: ratings are linked to lower-dimensional document-topic structure, rather than to isolated word-level effects.

The data-generating mechanism can be summarised as
\[
\mathbf{W}_{\mathrm{true}},\mathbf{H}_{\mathrm{true}}
\longrightarrow
\mathbf{X},
\qquad
\mathbf{W}_{\mathrm{true}},\boldsymbol{\beta}_{\mathrm{true}}
\longrightarrow
\mathbf{Y}.
\]
Thus, both the document-term matrix and the rating outcome share the same latent document-topic structure, while the rating is not generated directly from individual word counts.

In each simulated dataset, a true topic-word matrix
\(\mathbf{H}_{\mathrm{true}}\) and a true document-topic matrix \(\mathbf{W}_{\mathrm{true}}\) are first generated from Dirichlet distributions. The observed document-term matrix \(\mathbf{X}\) is then sampled as sparse word counts from the document-specific word probabilities implied by \(\mathbf{W}_{\mathrm{true}}\mathbf{H}_{\mathrm{true}}\). The rating outcome is generated separately from the latent topic proportions:
\[
\eta_d=\mathbf{w}_d^\top\boldsymbol{\beta}_{\mathrm{true}},
\qquad
p_d=\mathrm{logit}^{-1}(\eta_d),
\qquad
Y_d \sim \mathrm{Binomial}(4,p_d).
\]
The shifted outcome \(Y_d\in\{0,1,2,3,4\}\) corresponds to a five-level rating scale after adding one unit for presentation.

The baseline setting uses \(K_{\mathrm{true}}=5\), \(N=5000\), and an average document length of \(\bar{L}=30\), producing sparse short texts with clearly structured topics and a moderate-to-strong topic-level rating signal. The full data-generating mechanism and parameter settings are reported in Appendix~\ref{app:simulation_details}. The purpose of the simulation is therefore not to show that JBNMF dominates all predictive methods in general but to verify that it can recover rating-relevant latent structure when the topic-driven rating assumption holds.

\subsection{Evaluation metrics}\label{sec:evaluation-metrics}

Predictive performance is evaluated using \(R^2\) and root mean squared error (RMSE). For test observations \(i=1,\ldots,N_{\mathrm{test}}\),
\[
R^2
=
1-
\frac{\sum_{i=1}^{N_{\mathrm{test}}}(y_i-\hat{y}_i)^2}
{\sum_{i=1}^{N_{\mathrm{test}}}(y_i-\bar{y})^2},
\qquad
\mathrm{RMSE}
=
\sqrt{
\frac{1}{N_{\mathrm{test}}}
\sum_{i=1}^{N_{\mathrm{test}}}(y_i-\hat{y}_i)^2
}.
\]
Here, \(y_i\) is the observed rating, \(\hat{y}_i\) is the predicted rating, and \(\bar{y}\) is the mean observed rating in the test set. The test-set \(R^2\) measures improvement over a constant-mean predictor and is not constrained to be non-negative. A negative value therefore indicates that the model predicts worse than the test-set mean predictor. RMSE reports the absolute prediction error on the rating scale.
\subsection{Parameter sensitivity}\label{sec:sim-parameter-sensitivity}

We examine the sensitivity of JBNMF to the fitted number of topics \(K\) and the supervision weight \(\alpha\). These two parameters control different aspects of the model: \(K\) determines the dimension of the latent representation, while \(\alpha\) controls the relative contribution of the rating-learning component. Other numerical and stabilising parameters are fixed in this experiment, with implementation details reported in Appendix~\ref{app:simulation_details}.

Figure~\ref{fig:sim-sensitivity} reports test \(R^2\) and RMSE over
\(K=2,\ldots,15\) and
\(\alpha\in\{0.01,0.03,0.05,0.07,0.10\}\). The clearest pattern is that
the best performance occurs at \(K=5\), matching the true number of
generating topics. At this value, the model achieves \(R^2=0.704\) and
RMSE \(=0.795\), with little variation across the tested values of
\(\alpha\). This suggests that, when the fitted topic dimension matches the
data-generating structure, the recovered representation is stable within
this range of supervision weights.

The results also show that the effect of \(\alpha\) depends on the fitted
topic dimension. When \(K\) is too small, the model performs poorly across
all supervision weights: for example, \(K=2\) gives \(R^2\approx0.046\) and
RMSE \(=1.427\). When \(K\) is larger than the true value, performance can
remain reasonable for nearby choices, but it becomes more sensitive to
\(\alpha\). For instance, at \(K=6\), increasing \(\alpha\) from 0.01 to
0.10 reduces \(R^2\) from 0.533 to 0.196 and increases RMSE from 0.998 to
1.310. Similar deterioration appears for several over-specified models
under larger supervision weights.

Overall, the sensitivity analysis suggests an interaction between the
chosen topic dimension and the supervision weight. Correctly specifying, or
closely approximating, the latent topic dimension leads to stable
prediction, whereas stronger supervision can be harmful when the fitted
representation is over-specified. This supports tuning \(K\) and
\(\alpha\) jointly in the empirical analysis, while fixing numerical
stabilisation parameters to reduce the search space.

\begin{figure*}[!t]
\centering
\begin{minipage}{0.49\textwidth}
\centering
\includegraphics[width=\linewidth]{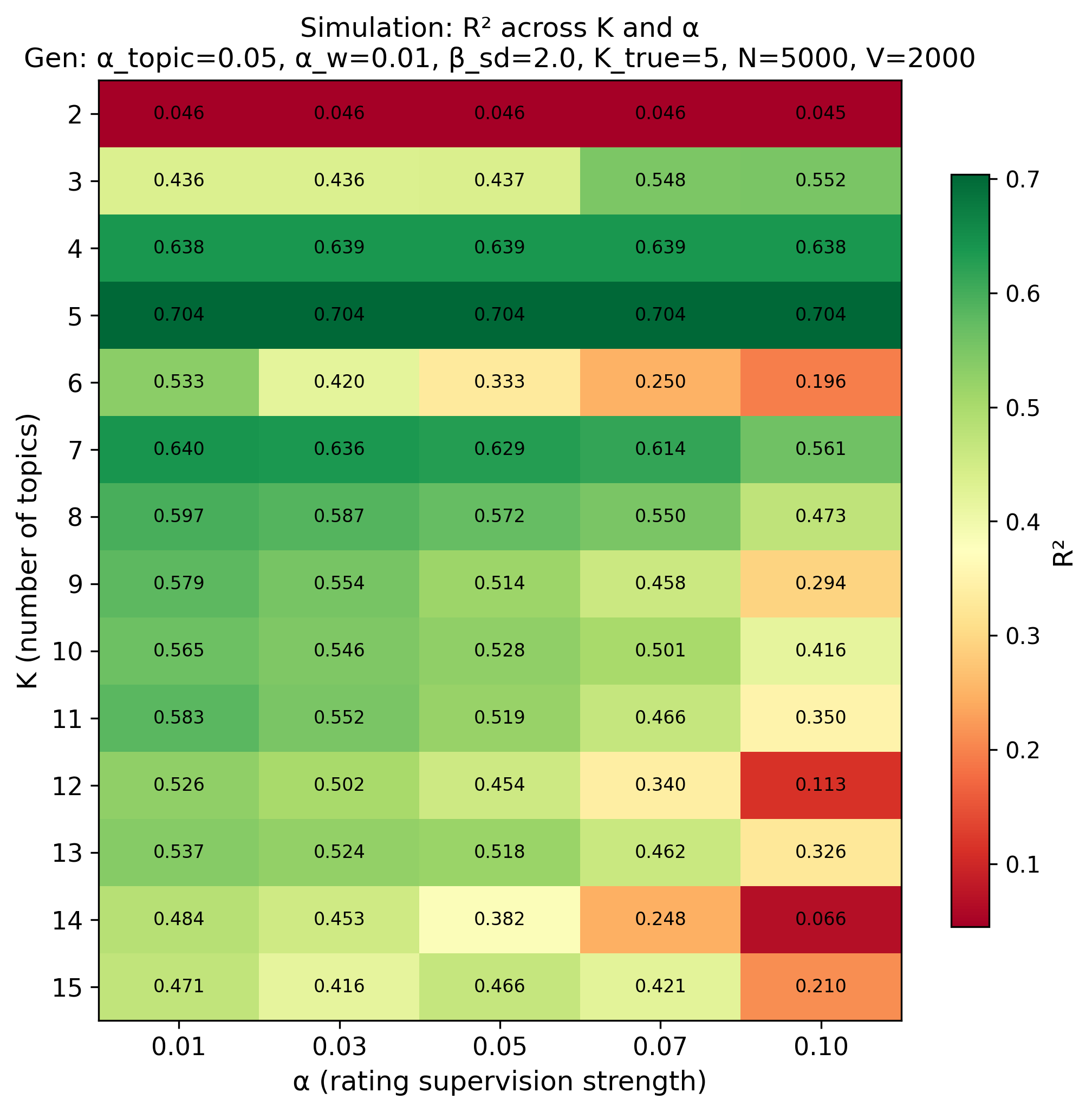}
\textbf{(a)} Test \(R^2\)
\end{minipage}
\hfill
\begin{minipage}{0.49\textwidth}
\centering
\includegraphics[width=\linewidth]{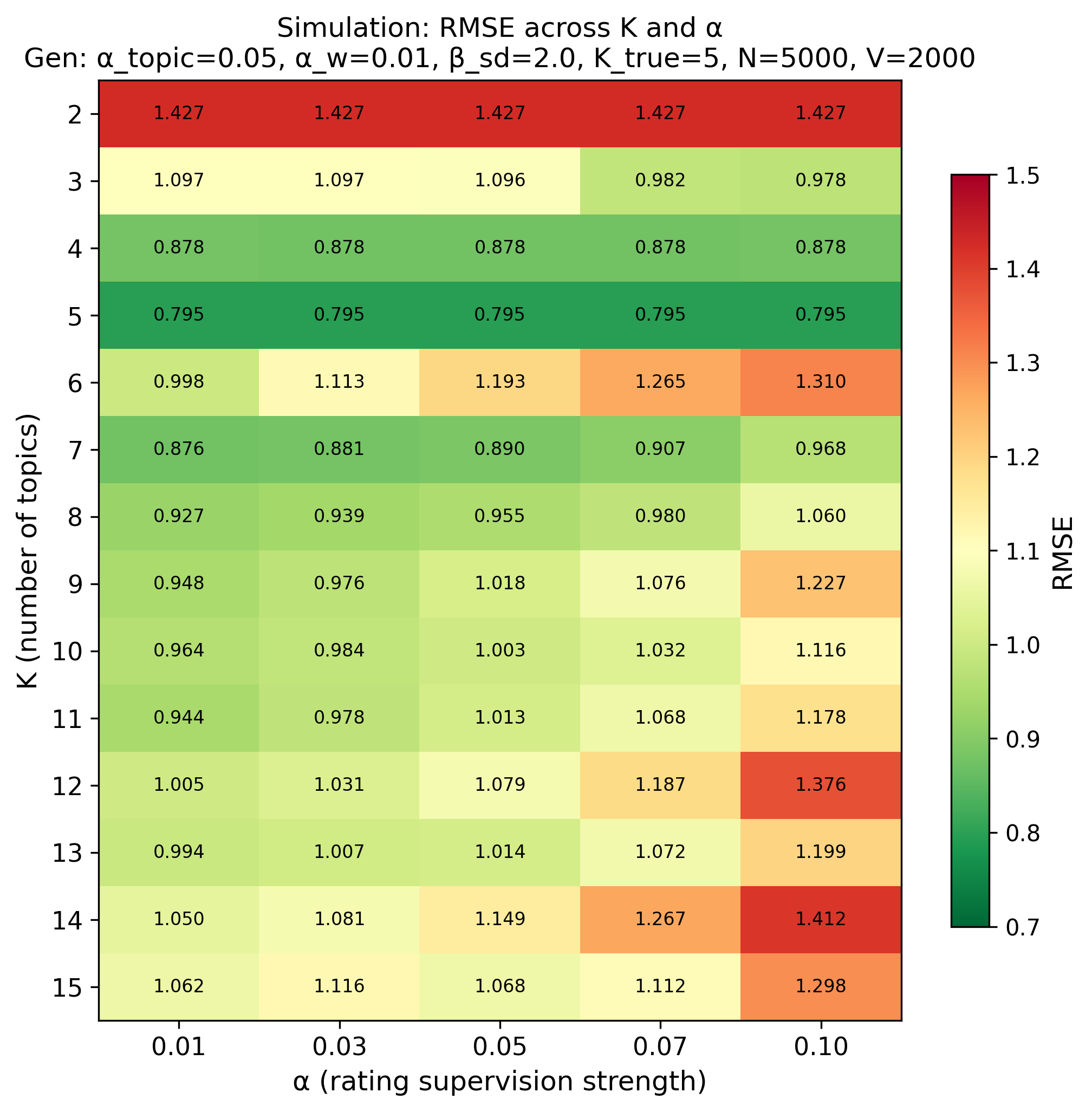}
\textbf{(b)} Test RMSE
\end{minipage}
\caption{Sensitivity of JBNMF prediction performance to the fitted number of topics \(K\) and the supervision weight \(\alpha\) in the simulation study. The data are generated with \(K_{\mathrm{true}}=5\), \(N=5000\), \(V=2000\), \(\alpha_{\mathrm{topic}}=0.05\), \(\alpha_w=0.01\), and \(\sigma_\beta=2.0\). Panel (a) reports test \(R^2\), and panel (b) reports test RMSE.\label{fig:sim-sensitivity}}%
\end{figure*}

\subsection{Model comparison}\label{sec:sim-model-comparison}

We next compare JBNMF with three predictive baselines: the mean predictor, unregularised linear regression and ridge regression. The linear and ridge models are fitted directly on the document-term representation, while JBNMF first estimates a low-dimensional topic representation and then links this representation to the rating outcome.

\begin{figure}[!t]
\centering
\includegraphics[width=\columnwidth]{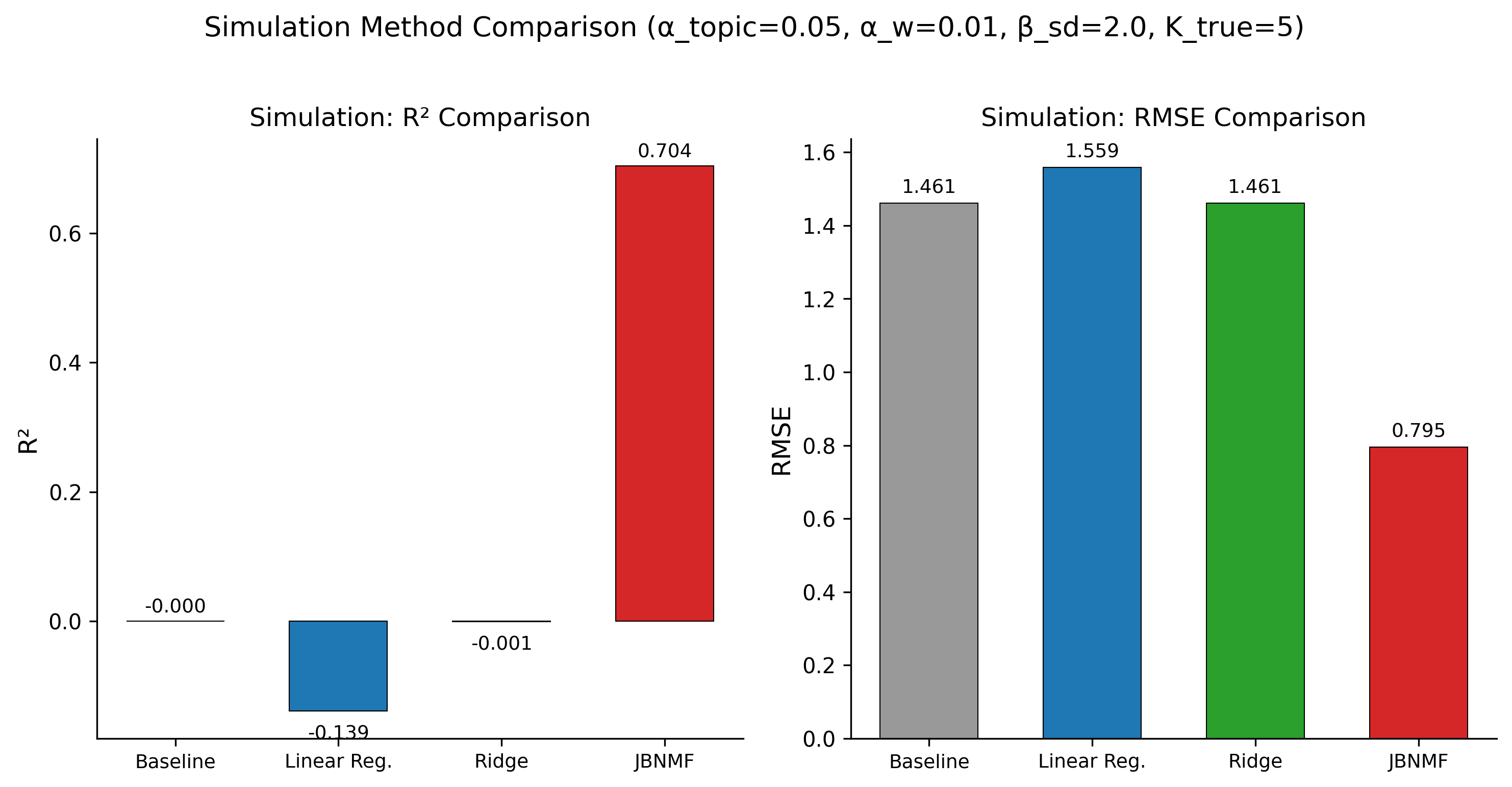}
\caption{Comparison between JBNMF and direct word-level predictive baselines in the simulation. The left panel reports test \(R^2\), and the right panel reports test RMSE. The data are generated from a topic-driven rating mechanism with \(K_{\mathrm{true}}=5\), \(\alpha_{\mathrm{topic}}=0.05\), \(\alpha_w=0.01\), and \(\sigma_\beta=2.0\).\label{fig:sim-method-comparison}}%
\end{figure}

Figure~\ref{fig:sim-method-comparison} shows a clear separation between the proposed model and the word-level baselines. The mean baseline gives \(R^2=0.000\) and RMSE \(=1.461\). Linear regression performs worse than the mean predictor, with \(R^2=-0.139\) and RMSE \(=1.559\), while ridge regression is effectively shrunk back to the baseline, with \(R^2=-0.001\) and RMSE \(=1.461\). By contrast, JBNMF achieves \(R^2=0.704\) and RMSE \(=0.795\).

This pattern follows directly from the simulation design. The rating outcome is not generated from isolated word-level effects. Instead, both the observed text and the rating outcome are driven by the same latent document-topic proportions:
\[
\mathbf{W}_{\mathrm{true}},\mathbf{H}_{\mathrm{true}}
\longrightarrow
\mathbf{X},
\qquad
\mathbf{W}_{\mathrm{true}},\boldsymbol{\beta}_{\mathrm{true}}
\longrightarrow
Y.
\]
The document-term matrix \(\mathbf{X}\) is therefore only a sparse and noisy textual realisation of the latent topic structure, generated under finite document length. With an average document length of 30 and a vocabulary size of 2000, the rating signal is dispersed across groups of topic-related words rather than concentrated in a small number of individual terms.

Under this mechanism, direct word-level regression is structurally mismatched to the data-generating process. It attempts to estimate the rating signal from high-dimensional sparse lexical features, while the true signal operates through the lower-dimensional topic mixture. JBNMF is better aligned with the simulation because it explicitly reconstructs this intermediate topic representation before using it for prediction. The simulation therefore confirms the central modelling assumption of the paper: when ratings are genuinely driven by latent semantic structure, a joint topic-rating model can recover and use this structure more effectively than direct word-level predictive baselines.

\section{Empirical application}\label{sec:application}

\subsection{Data description and empirical setup}

The simulation study provides a controlled benchmark in which the rating signal is known to be generated from latent topic proportions. The empirical analysis is more demanding: the topic-rating mechanism is no longer guaranteed, ratings are highly unevenly distributed, and review texts vary substantially in length and specificity. The purpose of the real-data analysis is therefore to examine when the proposed topic-based representation remains useful for prediction and interpretation outside the data-generating assumptions of the simulation.

We apply JBNMF to two real review data sources. The first is a TripAdvisor hotel review dataset obtained from Kaggle, where each observation consists of a free-text review and an associated five-point rating \citep{dhar_tripadvisor_20k}. The second is drawn from the Google Local Data (2021), restricted to New York businesses from selected service categories. The Google Local dataset contains review information, including ratings and review text, together with business metadata such as addresses, categories and geographical information \citep{zhang2021googlelocal,li2022uctopic,yan2023personalized}.
Using both sources allows us to assess the method across different review environments, including differences in rating behaviour, vocabulary use, review length and business context.

The main TripAdvisor sample contains \(N=10{,}000\) reviews. The rating distribution is concentrated at the upper end of the scale: 4-star and 5-star reviews account for 73.7\% of observations. This differs from the controlled simulation setting and reflects a common feature of online review data, where the prediction task often involves distinguishing degrees of satisfaction within a compressed outcome range.

Unless otherwise stated, the text preprocessing and TF-IDF representation follow the procedure described in Section~\ref{methodology}. The maximum raw feature size is set to 2000, and the final document-term matrix is restricted to 500 terms after TF-IDF filtering. For JBNMF, we fix \(\gamma=400\), \(\lambda=0.1\), the maximum number of iterations at 500 and the convergence tolerance at \(10^{-4}\). Each dataset is split into training and test sets using an 80/20 split. Predictive performance is evaluated using the \(R^2\) and RMSE metrics defined in Section~\ref{sec:evaluation-metrics}.

The empirical tuning procedure focuses on the fitted number of topics \(K\) and the supervision weight \(\alpha\), following the reduced search strategy motivated by the simulation study. We use a staged grid search: a coarse search is first used to locate a promising range of \(K\), followed by a denser local search around that region. The parameter \(\alpha\) is searched over small to moderate values to assess the contribution of the supervised rating component without allowing it to dominate the text reconstruction objective.
\subsection{Predictive performance under parameter search}
\label{sec:empirical-predictive-search}

We first examine the predictive performance of JBNMF on the main TripAdvisor sample under the reduced search over the fitted number of topics \(K\) and the supervision weight \(\alpha\). In the empirical setting, the true topic number is unknown, so the purpose of the search is to identify a well-performing configuration for subsequent comparison with predictive baselines. The full search results are reported in Appendix~\ref{app:empirical_parameter_search}.

Under this reduced search, the best result is obtained at \(K=14\) and \(\alpha=0.05\), giving \(R^2=0.551\) and RMSE \(=0.830\). This selected configuration is used in the main TripAdvisor comparison below.

To place the selected JBNMF configuration in context,
Table~\ref{tab:tripadvisor-main-performance} compares it with three predictive baselines on the same TripAdvisor test set. The comparison includes a mean-rating baseline, unregularised linear regression, and ridge regression fitted directly to the word-level representation.

\begin{table}[!t]
\caption{Predictive performance on the TripAdvisor review sample.\label{tab:tripadvisor-main-performance}}%
\tabcolsep=0pt%
\begin{tabular*}{\columnwidth}{@{\extracolsep{\fill}}lcc@{\extracolsep{\fill}}}
\toprule
Method & \(R^2\) & RMSE \\
\midrule
Mean baseline & 0.000 & 1.239 \\
Linear regression & 0.564 & 0.816 \\
Ridge regression & 0.569 & 0.812 \\
JBNMF \((K=14,\alpha=0.05)\) & 0.551 & 0.830 \\
\bottomrule
\end{tabular*}
\begin{tablenotes}%
\item Note: The JBNMF result corresponds to the best configuration from the reduced \(K,\alpha\) search. The mean baseline predicts the training-set mean rating.
\end{tablenotes}
\end{table}

The selected JBNMF model is slightly below the best word-level baseline in
pure predictive performance, but remains close to both linear and ridge
regression on this main TripAdvisor sample. The following sections examine
whether this level of predictive performance remains stable under changes in
sample size and whether the fitted topic structure provides useful semantic
information.
\subsection{Topic-level interpretation}
\label{sec:empirical-topic-interpretation}

The preceding results in Section~\ref{sec:empirical-predictive-search} show that JBNMF achieves predictive performance close to the word-level baselines on the main TripAdvisor sample. We now examine what is gained by using the topic-based representation. The fitted model provides two complementary quantities for interpretation: the regression coefficient \(\beta_k\), which measures the association between topic \(k\) and the rating outcome, and the document-topic weights \(W_{ik}\), which measure how strongly topic \(k\) appears in each review.

\begin{figure*}[!t]%
\centering
\includegraphics[width=\textwidth]{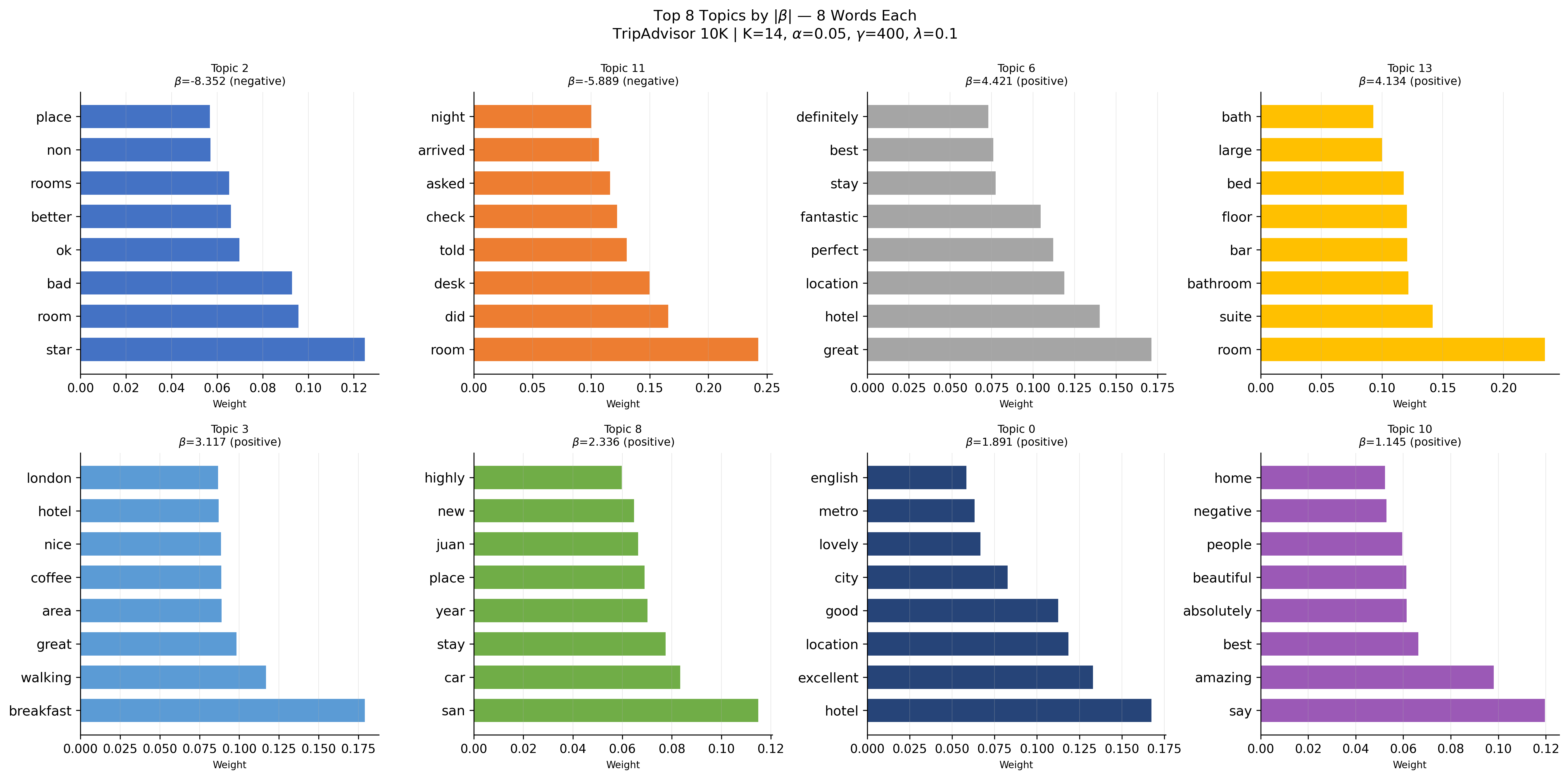}
\caption{Top eight TripAdvisor topics ranked by \(|\beta_k|\), where \(\beta_k\) measures the association between topic \(k\) and the rating outcome.\label{fig:top-beta-topics}}%
\end{figure*}

Figure~\ref{fig:top-beta-topics} ranks the eight most rating-sensitive topics by \(|\beta_k|\). This ranking highlights topics that are most strongly associated with rating variation, regardless of how frequently they occur in the corpus. The most negative topics are associated with terms describing poor room conditions, unpleasant smells, sleep disruption, front-desk interactions, charges and complaint-related language. In contrast, the most positive topics contain words related to favourable overall experience, location, service, breakfast, walking distance and premium room features such as suites, bathrooms and balconies. This suggests that the fitted topics are not merely collections of frequent words, but semantic dimensions that correspond to recognisable aspects of hotel experience.

\begin{figure*}[!t]%
\centering
\includegraphics[width=\textwidth]{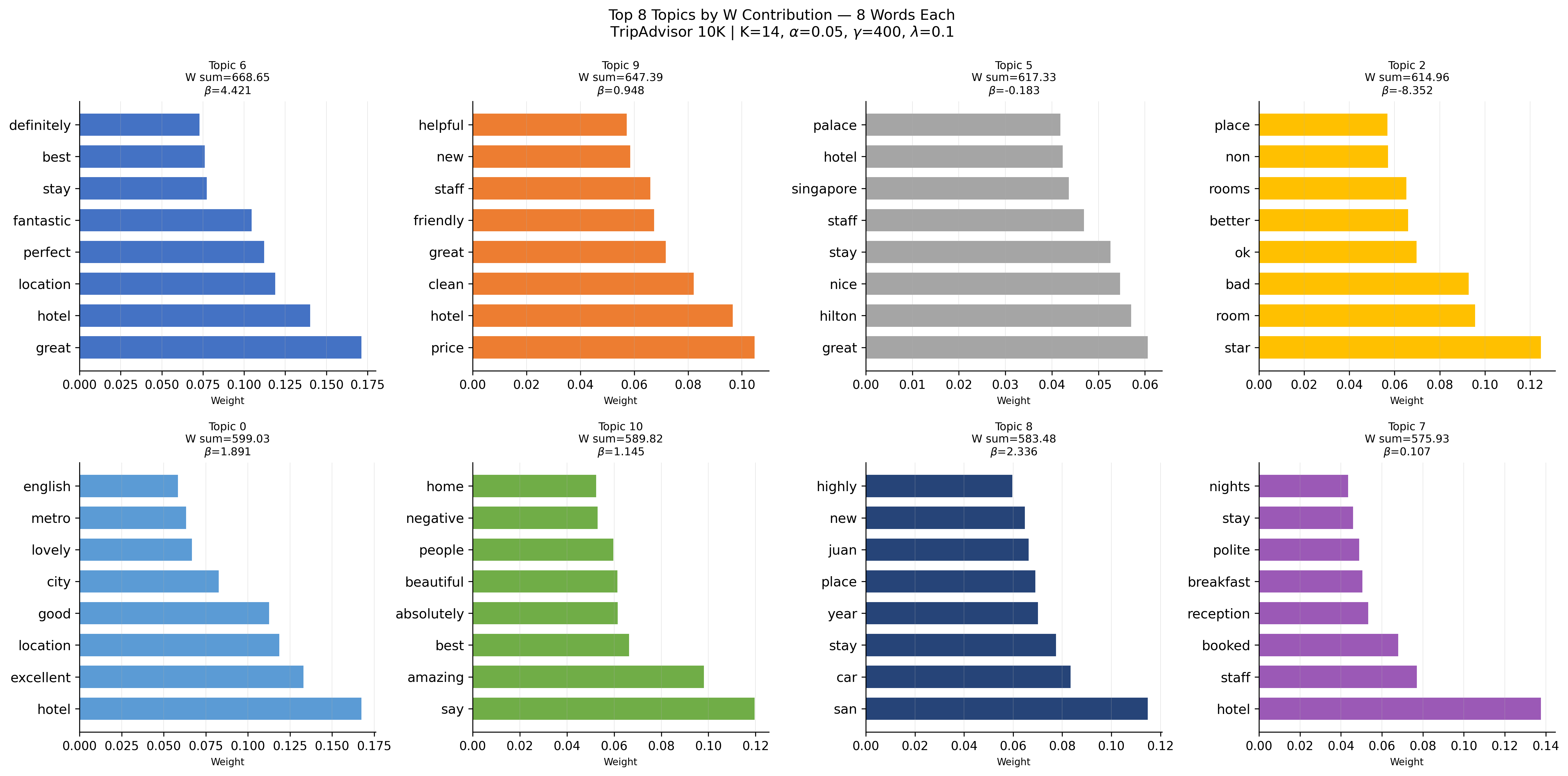}
\caption{Top eight TripAdvisor topics ranked by aggregate topic weight \(\sum_i W_{ik}\), which measures overall topic prevalence in the review corpus.\label{fig:top-w-topics}}%
\end{figure*}

Figure~\ref{fig:top-w-topics} gives a different view by ranking topics according to their aggregate document-topic weight \(\sum_i W_{ik}\). This measures prevalence rather than rating relevance. Some topics appear important under both rankings. For example, topics related to overall stay experience, location, room quality and service appear frequently in the corpus and also have clear rating associations. Other topics are common but have relatively small coefficients, meaning that they form part of the general background of hotel reviews without being especially discriminative for ratings. Conversely, a topic may have a large \(|\beta_k|\) even if it is not among the most prevalent themes, indicating a lower-frequency but high-impact experience dimension.

The distinction between \(|\beta_k|\) and \(\sum_i W_{ik}\) is the main interpretive benefit of the fitted model. The former identifies themes most strongly associated with rating changes, while the latter identifies themes most commonly discussed by reviewers. For business use, these are different questions. Topic prevalence indicates what customers talk about most often; topic coefficients indicate which themes are most closely linked to satisfaction or dissatisfaction. Taken together, the two rankings allow the fitted model to separate routine review content from rating-relevant experience dimensions, giving a more structured interpretation than a list of individual word coefficients.

Overall, the topic-level results indicate that the interpretability gained from JBNMF is substantively meaningful rather than merely formal. Although the model gives slightly lower predictive accuracy than the strongest word-level baseline on the main TripAdvisor sample, it produces rating-related topics that correspond to recognisable dimensions of hotel experience, including room quality, service, location, stay experience and complaint-related issues. This supports the central empirical trade-off of the model: a modest loss in pure predictive performance can be acceptable when it yields a structured semantic representation of the factors associated with ratings. Having established this trade-off, we next examine the conditions under which it is most useful.
\subsection{Sample size and overfitting}
\label{sec:empirical-sample-overfitting}

We next examine whether the predictive behaviour changes with sample size and whether JBNMF avoids the severe overfitting that can arise in unregularised word-level regression. For each method, we compare training and test \(R^2\), using the same definition as in Section~\ref{sec:evaluation-metrics}. We summarise overfitting by the train-test gap
\[
\Delta R^2
=
R^2_{\mathrm{train}}-R^2_{\mathrm{test}}.
\]
For JBNMF, training \(R^2\) is computed from the fitted training representation \(\mathbf{W}_{\mathrm{train}}\), while test \(R^2\) is computed after inferring \(\mathbf{W}_{\mathrm{test}}\) from the held-out text matrix with \(\mathbf{H}\) fixed.

\begin{figure*}[!t]
\centering
\includegraphics[width=\textwidth]{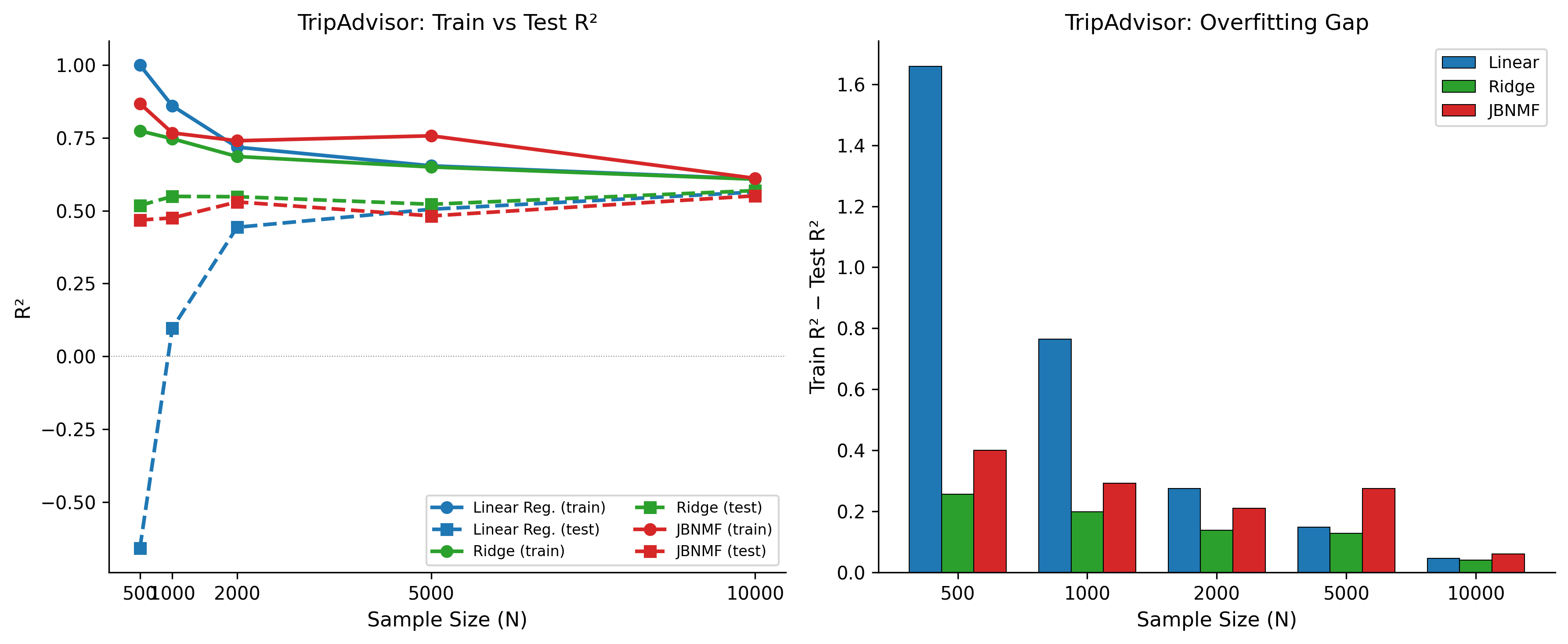}
\caption{Training and test \(R^2\) comparison on the TripAdvisor sample
across different sample sizes. The right panel reports the overfitting gap
\(\Delta R^2=R^2_{\mathrm{train}}-R^2_{\mathrm{test}}\).
\label{fig:trip-overfitting}}%
\end{figure*}

Figure~\ref{fig:trip-overfitting} reports the train-test comparison on the TripAdvisor data across different sample sizes. At \(M=500\), unregularised linear regression fits the training data almost perfectly but gives negative test \(R^2\), indicating severe overfitting. Ridge regression substantially reduces this gap. JBNMF also avoids the small-sample collapse of unregularised linear regression, with positive test performance and a much smaller train-test gap. As the sample size increases, the gap for linear regression narrows and the methods become closer in test performance.

Overall, JBNMF is not uniformly less overfitted than ridge regression, but it avoids the severe small-sample failure of unregularised word-level regression. This supports its role as a structured alternative when sample size is limited and direct high-dimensional word regression is unstable. A corresponding robustness check on a Google Local Hotel subset is reported in Appendix~\ref{app:additional-overfitting}.
\subsection{Review length and model applicability}
\label{sec:empirical-review-length}

We finally examine whether the usefulness of JBNMF depends on the amount of textual information available in each review. Since JBNMF estimates ratings through a document-topic representation, very short reviews may contain too little lexical evidence for stable topic recovery, even when word-level prediction remains possible.

\begin{figure*}[!t]
\centering
\includegraphics[width=\textwidth]{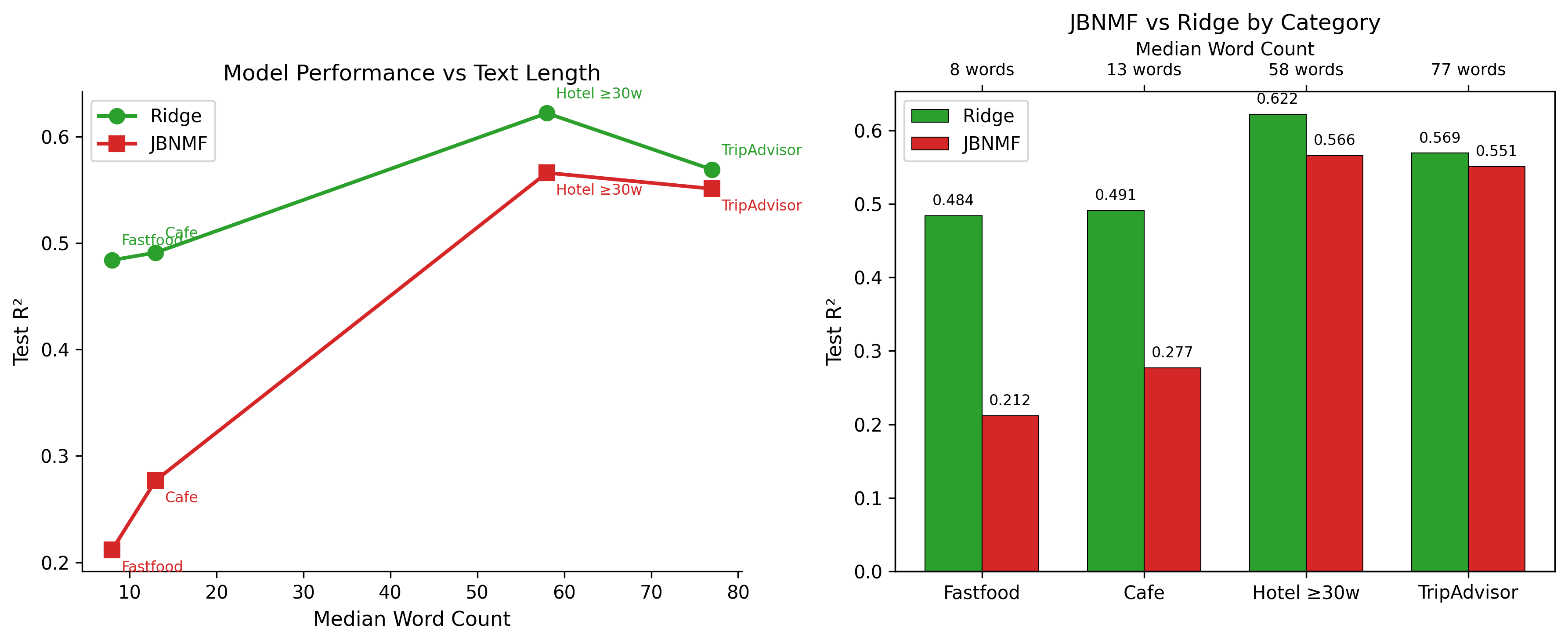}
\caption{Test \(R^2\) of JBNMF and ridge regression across review subsets with different median word counts. JBNMF is less competitive for very short reviews but approaches ridge regression when reviews contain more textual information.\label{fig:text-length-performance}}%
\end{figure*}

Figure~\ref{fig:text-length-performance} compares JBNMF and ridge regression across four review subsets with different median word counts. The TripAdvisor sample is the main empirical dataset. The remaining subsets are constructed from Google Local reviews for New York businesses in selected service categories: fast food, cafe, and hotel. Cafe and fast-food reviews are left unfiltered because they represent short-review settings, with median lengths of 13 and 8 words, respectively. For the Google Local Hotel subset, we retain reviews with at least 30 words to form a longer-review comparison group rather than to tune predictive performance; this gives a median length of 58 words and makes the subset closer to the TripAdvisor sample, whose median length is 77 words.

The results show a consistent dependence on review length. In the shortest settings, JBNMF is substantially below ridge regression: the \(R^2\) gap is 0.272 for fast-food reviews and 0.214 for cafe reviews. The gap narrows for longer reviews, falling to 0.056 for the filtered Google Local Hotel subset and 0.018 for the TripAdvisor sample. This comparison should not be read as a causal isolation of text length from category or platform effects, but it identifies a practical boundary: JBNMF becomes competitive when reviews contain enough textual information to support stable topic estimation.

Overall, JBNMF is not designed for extremely short, phrase-level reviews. Its value is strongest when the text is sufficiently informative and when a small loss in predictive accuracy is acceptable in exchange for interpretable topic-level structure.

\section{Discussion}\label{sec:discussion}

This paper has proposed JBNMF as an interpretable text-response modelling framework, using bounded star ratings as an initial case. The simulation confirms the intended mechanism: when the response is generated from latent topic mixtures, the model can recover response-relevant structure and outperform direct word-level regression. The TripAdvisor application gives a more realistic picture. JBNMF is slightly below ridge regression in pure predictive accuracy, with \(R^2=0.551\) and RMSE \(=0.830\), compared with \(R^2=0.569\) and RMSE \(=0.812\) for ridge regression. This result should not be interpreted as predictive superiority. Rather, it shows that a topic-mediated representation can remain competitive while providing a more inspectable semantic layer.

The main contribution is the prediction-interpretation trade-off. Ridge regression remains a strong benchmark when prediction alone is the objective, because it can exploit many regularised word-level associations. JBNMF provides a different output: by comparing topic-response coefficients with topic prevalence, it separates themes that are frequently discussed from themes that are more strongly associated with response variation. In the review application, these themes correspond to recognisable aspects of hotel experience, such as room quality, service, location, stay experience and complaint-related issues.

The empirical results also clarify when this trade-off is useful. JBNMF avoids the severe small-sample overfitting of unregularised word-level regression, although it is not uniformly less overfitted than ridge regression. It performs poorly for very short reviews, where there is too little lexical information to recover stable topic mixtures, but becomes more competitive for longer and more descriptive texts. The method is therefore most appropriate when text contains enough substantive information for topic estimation and when a small loss in predictive accuracy is acceptable in exchange for interpretable topic-level structure.

Several limitations remain. Topic interpretation depends on the quality of the recovered latent structure and some topics may remain mixed in real text data. The supervision weight \(\alpha\) is tied to the relative scale of the reconstruction and response losses, so its numerical value should not be compared directly across datasets. The current binomial formulation is also specific to bounded ratings. More generally, however, the proposed model can be viewed as an instance of a broader GLM-NMF framework, in which the document-topic vector \(\mathbf{w}_i\) provides interpretable covariates and \(\mathbf{w}_i^\top\boldsymbol{\beta}\) acts as the GLM linear predictor. Under this view, analysing other response types would mainly require replacing the binomial rating loss with the negative log-likelihood of an appropriate model, such as Gaussian, Bernoulli, Poisson or ordinal regression. Such extensions would move the method beyond online reviews towards broader applications in biomedical research, survey analysis, finance, education and risk assessment, where textual evidence and structured responses need to be analysed together.


\section{Competing interests}
No competing interest is declared.

\section{Author contributions statement}

C.J. led the study, including model development, algorithm implementation, simulation design, empirical analysis, result interpretation, and manuscript drafting. 
B.P. contributed to the optimisation strategy, model refinement, experimental supervision, result verification, troubleshooting, and manuscript editing. 
N.M. contributed to the mathematical formulation, verification of equations and update rules, experimental supervision, and manuscript editing. 
All authors reviewed the manuscript and approved the final version.

\section{Acknowledgments}

\bibliographystyle{plainnat}
\bibliography{reference}

\begin{thebibliography}{35}
\providecommand{\natexlab}[1]{#1}
\providecommand{\url}[1]{\texttt{#1}}
\expandafter\ifx\csname urlstyle\endcsname\relax
  \providecommand{\doi}[1]{doi: #1}\else
  \providecommand{\doi}{doi: \begingroup \urlstyle{rm}\Url}\fi

\bibitem[Agresti(2015)]{agresti2015foundations}
Alan Agresti.
\newblock \emph{Foundations of Linear and Generalized Linear Models}.
\newblock Wiley, Hoboken, NJ, 2015.

\bibitem[Archak et~al.(2011)Archak, Ghose, and Ipeirotis]{archak2011deriving}
Nikolay Archak, Anindya Ghose, and Panagiotis~G. Ipeirotis.
\newblock Deriving the pricing power of product features by mining consumer reviews.
\newblock \emph{Management Science}, 57\penalty0 (8):\penalty0 1485--1509, 2011.
\newblock \doi{10.1287/mnsc.1110.1370}.

\bibitem[Blei et~al.(2003)Blei, Ng, and Jordan]{blei2003latent}
David~M. Blei, Andrew~Y. Ng, and Michael~I. Jordan.
\newblock Latent dirichlet allocation.
\newblock \emph{Journal of Machine Learning Research}, 3:\penalty0 993--1022, 2003.
\newblock URL \url{https://jmlr.csail.mit.edu/papers/v3/blei03a.html}.

\bibitem[Chevalier and Mayzlin(2006)]{chevalier2006effect}
Judith~A. Chevalier and Dina Mayzlin.
\newblock The effect of word of mouth on sales: Online book reviews.
\newblock \emph{Journal of Marketing Research}, 43\penalty0 (3):\penalty0 345--354, 2006.
\newblock \doi{10.1509/jmkr.43.3.345}.

\bibitem[Devlin et~al.(2019)Devlin, Chang, Lee, and Toutanova]{devlin2019bert}
Jacob Devlin, Ming-Wei Chang, Kenton Lee, and Kristina Toutanova.
\newblock {BERT}: Pre-training of deep bidirectional transformers for language understanding.
\newblock In \emph{Proceedings of the 2019 Conference of the North American Chapter of the Association for Computational Linguistics: Human Language Technologies}, volume~1, pages 4171--4186. Association for Computational Linguistics, 2019.
\newblock \doi{10.18653/v1/N19-1423}.
\newblock URL \url{https://doi.org/10.18653/v1/N19-1423}.

\bibitem[Dhar(2020)]{dhar_tripadvisor_20k}
Shoumik Dhar.
\newblock Tripadvisor hotel reviews 20k dataset.
\newblock \url{https://www.kaggle.com/datasets/shoumikdhar/tripadvisor-hotel-reviews-20k-dataset}, 2020.
\newblock Accessed: 2026-05-28.

\bibitem[Griffiths and Steyvers(2004)]{griffiths2004finding}
Thomas~L. Griffiths and Mark Steyvers.
\newblock Finding scientific topics.
\newblock \emph{Proceedings of the National Academy of Sciences}, 101\penalty0 (suppl. 1):\penalty0 5228--5235, 2004.
\newblock \doi{10.1073/pnas.0307752101}.
\newblock URL \url{https://doi.org/10.1073/pnas.0307752101}.

\bibitem[Hastie et~al.(2009)Hastie, Tibshirani, and Friedman]{hastie2009elements}
Trevor Hastie, Robert Tibshirani, and Jerome Friedman.
\newblock \emph{The Elements of Statistical Learning: Data Mining, Inference, and Prediction}.
\newblock Springer, New York, 2 edition, 2009.
\newblock \doi{10.1007/978-0-387-84858-7}.
\newblock URL \url{https://doi.org/10.1007/978-0-387-84858-7}.

\bibitem[Hoerl and Kennard(1970)]{hoerl1970ridge}
Arthur~E. Hoerl and Robert~W. Kennard.
\newblock Ridge regression: Biased estimation for nonorthogonal problems.
\newblock \emph{Technometrics}, 12\penalty0 (1):\penalty0 55--67, 1970.
\newblock \doi{10.1080/00401706.1970.10488634}.
\newblock URL \url{https://doi.org/10.1080/00401706.1970.10488634}.

\bibitem[Hu et~al.(2009)Hu, Zhang, and Pavlou]{hu2009overcoming}
Nan Hu, Jie Zhang, and Paul~A. Pavlou.
\newblock Overcoming the {J}-shaped distribution of product reviews.
\newblock \emph{Communications of the ACM}, 52\penalty0 (10):\penalty0 144--147, 2009.
\newblock \doi{10.1145/1562764.1562800}.

\bibitem[Hutto and Gilbert(2014)]{hutto2014vader}
C.~J. Hutto and Eric Gilbert.
\newblock {VADER}: A parsimonious rule-based model for sentiment analysis of social media text.
\newblock In \emph{Proceedings of the International {AAAI} Conference on Web and Social Media}, pages 216--225, 2014.
\newblock \doi{10.1609/icwsm.v8i1.14550}.

\bibitem[Jacovi and Goldberg(2020)]{jacovi2020towards}
Alon Jacovi and Yoav Goldberg.
\newblock Towards faithfully interpretable {NLP} systems: How should we define and evaluate faithfulness?
\newblock In \emph{Proceedings of the 58th Annual Meeting of the Association for Computational Linguistics}, pages 4198--4205. Association for Computational Linguistics, 2020.
\newblock \doi{10.18653/v1/2020.acl-main.386}.
\newblock URL \url{https://doi.org/10.18653/v1/2020.acl-main.386}.

\bibitem[Kim et~al.(2014)Kim, He, and Park]{kim2014algorithms}
Jingu Kim, Yunlong He, and Haesun Park.
\newblock Algorithms for nonnegative matrix and tensor factorizations: A unified view based on block coordinate descent framework.
\newblock \emph{Journal of Global Optimization}, 58\penalty0 (2):\penalty0 285--319, 2014.
\newblock \doi{10.1007/s10898-013-0035-4}.

\bibitem[Lee and Seung(1999)]{lee1999learning}
Daniel~D. Lee and H.~Sebastian Seung.
\newblock Learning the parts of objects by non-negative matrix factorization.
\newblock \emph{Nature}, 401\penalty0 (6755):\penalty0 788--791, 1999.
\newblock \doi{10.1038/44565}.

\bibitem[Lee and Seung(2001)]{lee2001algorithms}
Daniel~D. Lee and H.~Sebastian Seung.
\newblock Algorithms for non-negative matrix factorization.
\newblock In \emph{Advances in Neural Information Processing Systems 13}, pages 556--562. MIT Press, 2001.

\bibitem[Li et~al.(2022)Li, Shang, and McAuley]{li2022uctopic}
Jiacheng Li, Jingbo Shang, and Julian McAuley.
\newblock {UCT}opic: Unsupervised contrastive learning for phrase representations and topic mining.
\newblock In \emph{Proceedings of the 60th Annual Meeting of the Association for Computational Linguistics (Volume 1: Long Papers)}, pages 6159--6169, Dublin, Ireland, 2022. Association for Computational Linguistics.
\newblock \doi{10.18653/v1/2022.acl-long.426}.
\newblock URL \url{https://aclanthology.org/2022.acl-long.426/}.

\bibitem[Lipton(2018)]{lipton2018mythos}
Zachary~C. Lipton.
\newblock The mythos of model interpretability.
\newblock \emph{Communications of the ACM}, 61\penalty0 (10):\penalty0 36--43, 2018.
\newblock \doi{10.1145/3233231}.
\newblock URL \url{https://doi.org/10.1145/3233231}.

\bibitem[Liu(2012)]{liu2012sentiment}
Bing Liu.
\newblock \emph{Sentiment Analysis and Opinion Mining}.
\newblock Synthesis Lectures on Human Language Technologies. Morgan \& Claypool Publishers, San Rafael, CA, 2012.
\newblock \doi{10.2200/S00416ED1V01Y201204HLT016}.
\newblock URL \url{https://doi.org/10.2200/S00416ED1V01Y201204HLT016}.

\bibitem[Lu et~al.(2011)Lu, Ott, Cardie, and Tsou]{lu2011multi}
Bin Lu, Myle Ott, Claire Cardie, and Benjamin~K. Tsou.
\newblock Multi-aspect sentiment analysis with topic models.
\newblock In \emph{Proceedings of the 2011 {IEEE} 11th International Conference on Data Mining Workshops}, pages 81--88. {IEEE}, 2011.
\newblock \doi{10.1109/ICDMW.2011.125}.

\bibitem[Luca(2016)]{luca2016reviews}
Michael Luca.
\newblock Reviews, reputation, and revenue: The case of {Yelp.com}.
\newblock Working Paper 12-016, Harvard Business School, 2016.
\newblock Revised March 2016.

\bibitem[McAuliffe and Blei(2008)]{mcauliffe2008supervised}
Jon~D. McAuliffe and David~M. Blei.
\newblock Supervised topic models.
\newblock In \emph{Advances in Neural Information Processing Systems}, volume~20, pages 121--128, 2008.

\bibitem[McCullagh and Nelder(1989)]{mccullagh1989generalized}
P.~McCullagh and J.~A. Nelder.
\newblock \emph{Generalized Linear Models}.
\newblock Chapman \& Hall, London, 2 edition, 1989.

\bibitem[McCullagh(1980)]{mccullagh1980regression}
Peter McCullagh.
\newblock Regression models for ordinal data.
\newblock \emph{Journal of the Royal Statistical Society: Series B (Methodological)}, 42\penalty0 (2):\penalty0 109--142, 1980.
\newblock \doi{10.1111/j.2517-6161.1980.tb01109.x}.

\bibitem[Pang and Lee(2008)]{pang2008opinion}
Bo~Pang and Lillian Lee.
\newblock Opinion mining and sentiment analysis.
\newblock \emph{Foundations and Trends in Information Retrieval}, 2\penalty0 (1--2):\penalty0 1--135, 2008.
\newblock \doi{10.1561/1500000011}.

\bibitem[Pennebaker et~al.(2015)Pennebaker, Booth, Boyd, and Francis]{pennebaker2015liwc}
James~W. Pennebaker, Roger~J. Booth, Ryan~L. Boyd, and Martha~E. Francis.
\newblock \emph{Linguistic Inquiry and Word Count: {LIWC2015}}.
\newblock Pennebaker Conglomerates, Austin, TX, 2015.
\newblock URL \url{https://www.liwc.app/}.

\bibitem[Roberts et~al.(2014)Roberts, Stewart, Tingley, Lucas, Leder-Luis, Gadarian, Albertson, and Rand]{roberts2014structural}
Margaret~E. Roberts, Brandon~M. Stewart, Dustin Tingley, Christopher Lucas, Jetson Leder-Luis, Shana~Kushner Gadarian, Bethany Albertson, and David~G. Rand.
\newblock Structural topic models for open-ended survey responses.
\newblock \emph{American Journal of Political Science}, 58\penalty0 (4):\penalty0 1064--1082, 2014.
\newblock \doi{10.1111/ajps.12103}.

\bibitem[Salton and Buckley(1988)]{salton1988term}
Gerard Salton and Christopher Buckley.
\newblock Term-weighting approaches in automatic text retrieval.
\newblock \emph{Information Processing \& Management}, 24\penalty0 (5):\penalty0 513--523, 1988.
\newblock \doi{10.1016/0306-4573(88)90021-0}.
\newblock URL \url{https://doi.org/10.1016/0306-4573(88)90021-0}.

\bibitem[Taddy(2013)]{taddy2013multinomial}
Matt Taddy.
\newblock Multinomial inverse regression for text analysis.
\newblock \emph{Journal of the American Statistical Association}, 108\penalty0 (503):\penalty0 755--770, 2013.
\newblock \doi{10.1080/01621459.2012.734168}.

\bibitem[Tibshirani(1996)]{tibshirani1996regression}
Robert Tibshirani.
\newblock Regression shrinkage and selection via the lasso.
\newblock \emph{Journal of the Royal Statistical Society: Series B}, 58\penalty0 (1):\penalty0 267--288, 1996.
\newblock \doi{10.1111/j.2517-6161.1996.tb02080.x}.
\newblock URL \url{https://doi.org/10.1111/j.2517-6161.1996.tb02080.x}.

\bibitem[Titov and McDonald(2008)]{titov2008modeling}
Ivan Titov and Ryan~T. McDonald.
\newblock Modeling online reviews with multi-grain topic models.
\newblock In \emph{Proceedings of the 17th International Conference on World Wide Web}, pages 111--120. Association for Computing Machinery, 2008.
\newblock \doi{10.1145/1367497.1367513}.
\newblock URL \url{https://doi.org/10.1145/1367497.1367513}.

\bibitem[Vaswani et~al.(2017)Vaswani, Shazeer, Parmar, Uszkoreit, Jones, Gomez, Kaiser, and Polosukhin]{vaswani2017attention}
Ashish Vaswani, Noam Shazeer, Niki Parmar, Jakob Uszkoreit, Llion Jones, Aidan~N. Gomez, Lukasz Kaiser, and Illia Polosukhin.
\newblock Attention is all you need.
\newblock In \emph{Advances in Neural Information Processing Systems}, volume~30, pages 5998--6008, 2017.
\newblock URL \url{https://papers.nips.cc/paper/7181-attention-is-all-you-need}.

\bibitem[Xu et~al.(2003)Xu, Liu, and Gong]{xu2003document}
Wei Xu, Xin Liu, and Yihong Gong.
\newblock Document clustering based on non-negative matrix factorization.
\newblock In \emph{Proceedings of the 26th Annual International ACM SIGIR Conference on Research and Development in Information Retrieval}, pages 267--273. ACM, 2003.
\newblock \doi{10.1145/860435.860485}.

\bibitem[Yan et~al.(2023)Yan, He, Li, Zhang, and McAuley]{yan2023personalized}
An~Yan, Zhankui He, Jiacheng Li, Tianyang Zhang, and Julian McAuley.
\newblock Personalized showcases: Generating multi-modal explanations for recommendations.
\newblock In \emph{Proceedings of the 46th International ACM SIGIR Conference on Research and Development in Information Retrieval}, pages 2251--2255. Association for Computing Machinery, 2023.
\newblock \doi{10.1145/3539618.3592036}.
\newblock URL \url{https://doi.org/10.1145/3539618.3592036}.

\bibitem[Yan et~al.(2013)Yan, Guo, Lan, and Cheng]{yan2013biterm}
Xiaohui Yan, Jiafeng Guo, Yanyan Lan, and Xueqi Cheng.
\newblock A biterm topic model for short texts.
\newblock In \emph{Proceedings of the 22nd International Conference on World Wide Web}, pages 1445--1456. Association for Computing Machinery, 2013.
\newblock \doi{10.1145/2488388.2488514}.

\bibitem[Zhang and Li(2021)]{zhang2021googlelocal}
Tianyang Zhang and Jiacheng Li.
\newblock Google local data (2021).
\newblock \url{https://mcauleylab.ucsd.edu/public_datasets/gdrive/googlelocal/}, 2021.
\newblock Accessed: 2026-05-28.

\end{thebibliography}

\appendix

\numberwithin{equation}{section}
\numberwithin{figure}{section}
\numberwithin{table}{section}

\section{Derivation of the optimisation updates}
\label{app:update_derivations}

This section provides the derivation of the update rules used in the Joint Binomial NMF algorithm. The notation is consistent with the model formulation in the main text. The document-term matrix is denoted by \(\mathbf{X}\in\mathbb{R}^{M\times N}_{+}\), the document-topic matrix by \(\mathbf{W}\in\mathbb{R}^{M\times K}_{+}\), the topic-term matrix by \(\mathbf{H}\in\mathbb{R}^{K\times N}_{+}\), and the regression coefficient vector by \(\boldsymbol{\beta}\in\mathbb{R}^{K}\).

The full objective is
\begin{equation}
\mathcal{J}(\mathbf{W},\mathbf{H},\boldsymbol{\beta})
=
\mathcal{L}_{\mathrm{text}}(\mathbf{X},\mathbf{W}\mathbf{H})
+
\alpha \mathcal{L}_{\mathrm{rating}}(\mathbf{Y},\mathbf{W},\boldsymbol{\beta})
+
\lambda\Vert\boldsymbol{\beta}\Vert_2^2 ,
\label{app:eq:full_objective}
\end{equation}
where \(\alpha\geq 0\) controls the contribution of the rating loss and \(\lambda\geq 0\) controls the strength of the \(L_2\) penalty. The KL reconstruction loss is
\begin{equation}
\mathcal{L}_{\mathrm{text}}
=
\sum_{i=1}^{M}\sum_{j=1}^{N}
\left[
X_{ij}\log\frac{X_{ij}}{(\mathbf{W}\mathbf{H})_{ij}}
-
X_{ij}
+
(\mathbf{W}\mathbf{H})_{ij}
\right],
\label{app:eq:text_loss}
\end{equation}
with the usual convention that the term \(X_{ij}\log\{X_{ij}/(\mathbf{W}\mathbf{H})_{ij}\}\) is zero when \(X_{ij}=0\). For \(Y_i\in\{0,1,2,3,4\}\), the binomial regression component is
\begin{equation}
\mathcal{L}_{\mathrm{rating}}
=
-
\sum_{i=1}^{M}
\left[
Y_i\log p_i
+
(4-Y_i)\log(1-p_i)
\right],
\label{app:eq:rating_loss}
\end{equation}
where
\[
p_i=\sigma(\mathbf{W}_{i\cdot}\boldsymbol{\beta}),
\qquad
\sigma(z)=\frac{1}{1+\exp(-z)}.
\]

For a nonnegative parameter \(a\geq 0\), the multiplicative updates can be motivated from the KKT complementary slackness condition. If the relevant objective is denoted by \(\mathcal{J}(a)\), the KKT condition implies \(a\,\partial\mathcal{J}/\partial a=0\) at a stationary point, after accounting for the nonnegativity constraint. When the gradient can be written as the difference of positive and negative components,
\[
\frac{\partial\mathcal{J}}{\partial a}
=
[\nabla\mathcal{J}]_{+}
-
[\nabla\mathcal{J}]_{-},
\]
the fixed-point relation \(a[\nabla\mathcal{J}]_{+}=a[\nabla\mathcal{J}]_{-}\) motivates the multiplicative update
\[
a
\leftarrow
a
\frac{[\nabla\mathcal{J}]_{-}}
{[\nabla\mathcal{J}]_{+}} .
\]

\subsection{Update for the topic-term matrix}
\label{app:update_h}

Given \(\mathbf{W}\), the topic-term matrix \(\mathbf{H}\) appears only in the KL reconstruction loss. Differentiating Eq.~\eqref{app:eq:text_loss} with respect to \(H_{kj}\) gives
\begin{equation}
\frac{\partial \mathcal{L}_{\mathrm{text}}}{\partial H_{kj}}
=
\sum_{i=1}^{M} W_{ik}
-
\sum_{i=1}^{M}
W_{ik}
\frac{X_{ij}}{(\mathbf{W}\mathbf{H})_{ij}} .
\label{app:eq:grad_h}
\end{equation}
Applying the KKT complementary slackness condition for the constraint \(H_{kj}\geq 0\) gives \(H_{kj}\,\partial\mathcal{L}_{\mathrm{text}}/\partial H_{kj}=0\) at a stationary point. Separating the positive and negative parts of the gradient in Eq.~\eqref{app:eq:grad_h} gives the fixed-point relation
\begin{equation}
H_{kj}
\sum_{i=1}^{M} W_{ik}
=
H_{kj}
\sum_{i=1}^{M}
W_{ik}
\frac{X_{ij}}{(\mathbf{W}\mathbf{H})_{ij}} .
\label{app:eq:h_fixed_point}
\end{equation}
This yields the element-wise update
\begin{equation}
H_{kj}
\leftarrow
H_{kj}
\frac{
\sum_{i=1}^{M}
W_{ik}
\frac{X_{ij}}{(\mathbf{W}\mathbf{H})_{ij}}
}{
\sum_{i=1}^{M} W_{ik}
}.
\label{app:eq:h_update_element}
\end{equation}
In matrix form, this can be written as
\begin{equation}
\mathbf{H}
\leftarrow
\mathbf{H}
\odot
\frac{
\mathbf{W}^{\top}
\left[
\mathbf{X}/(\mathbf{W}\mathbf{H})
\right]
}{
\mathbf{W}^{\top}\mathbf{1}_{M}
},
\label{app:eq:h_update_matrix}
\end{equation}
where \(\odot\) denotes element-wise multiplication, the division is
element-wise, and \(\mathbf{1}_{M}\) is a column vector of ones of length
\(M\).

\subsection{Update for the regression coefficients}
\label{app:update_beta}

Conditional on \(\mathbf{W}\), the regression coefficients
\(\boldsymbol{\beta}\) appear in the binomial regression component and in
the \(L_2\) penalty. For each coefficient \(\beta_k\), the derivative of
the binomial loss is
\begin{equation}
\frac{\partial \mathcal{L}_{\mathrm{rating}}}{\partial \beta_k}
=
\sum_{i=1}^{M}
W_{ik}(4p_i-Y_i).
\label{app:eq:beta_grad_element}
\end{equation}
After adding the derivative of the \(L_2\) penalty, the full gradient is
\begin{equation}
\nabla_{\boldsymbol{\beta}}
\left(
\mathcal{L}_{\mathrm{rating}}
+
\lambda\Vert\boldsymbol{\beta}\Vert_2^2
\right)
=
\mathbf{W}^{\top}(4\mathbf{p}-\mathbf{Y})
+
2\lambda\boldsymbol{\beta},
\label{app:eq:beta_grad_matrix}
\end{equation}
where \(\mathbf{p}=(p_1,\ldots,p_M)^{\top}\) and
\(\mathbf{Y}=(Y_1,\ldots,Y_M)^{\top}\).

Since the regression part enters the full objective through the supervision
weight \(\alpha\), the gradient descent update for
\(\boldsymbol{\beta}\) is
\begin{equation}
\boldsymbol{\beta}
\leftarrow
\boldsymbol{\beta}
-
\eta\alpha
\left[
\mathbf{W}^{\top}(4\mathbf{p}-\mathbf{Y})
+
2\lambda\boldsymbol{\beta}
\right],
\label{app:eq:beta_update}
\end{equation}
where \(\eta>0\) is the learning rate. Unlike \(\mathbf{W}\) and
\(\mathbf{H}\), the vector \(\boldsymbol{\beta}\) is not constrained to be
nonnegative, since topics may have either positive or negative associations
with rating outcomes.

\subsection{Update for the document-topic matrix}
\label{app:update_w}

The document-topic matrix \(\mathbf{W}\) appears in both the KL
reconstruction loss and the binomial likelihood. The part of the objective
depending on \(\mathbf{W}\) is
\[
\mathcal{J}_{\mathbf{W}}
=
\mathcal{L}_{\mathrm{text}}
+
\alpha\mathcal{L}_{\mathrm{rating}},
\]
where the \(L_2\) penalty on \(\boldsymbol{\beta}\) does not contribute to
the derivative with respect to \(\mathbf{W}\).

The derivative of the KL reconstruction term with respect to \(W_{ik}\) is
\begin{equation}
\frac{\partial \mathcal{L}_{\mathrm{text}}}{\partial W_{ik}}
=
\sum_{j=1}^{N} H_{kj}
-
\sum_{j=1}^{N}
H_{kj}
\frac{X_{ij}}{(\mathbf{W}\mathbf{H})_{ij}} .
\label{app:eq:grad_w_reconstruction}
\end{equation}
For the binomial component, using
\(p_i=\sigma(\mathbf{W}_{i\cdot}\boldsymbol{\beta})\) gives
\begin{equation}
\frac{\partial \mathcal{L}_{\mathrm{rating}}}{\partial W_{ik}}
=
(4p_i-Y_i)\beta_k .
\label{app:eq:grad_w_binomial}
\end{equation}
Combining these terms, the derivative of \(\mathcal{J}_{\mathbf{W}}\) is
\begin{equation}
\frac{\partial \mathcal{J}_{\mathbf{W}}}{\partial W_{ik}}
=
\sum_{j=1}^{N} H_{kj}
-
\sum_{j=1}^{N}
H_{kj}
\frac{X_{ij}}{(\mathbf{W}\mathbf{H})_{ij}}
+
\alpha(4p_i-Y_i)\beta_k .
\label{app:eq:grad_w_joint}
\end{equation}

Applying the same KKT argument to the constraint \(W_{ik}\geq 0\) gives
\(W_{ik}\,\partial\mathcal{J}_{\mathbf{W}}/\partial W_{ik}=0\) at a
stationary point. To obtain a multiplicative update that preserves the
constraint \(W_{ik}\geq 0\), we decompose the gradient into positive and
negative parts. Define
\[
G_{ik}=(4p_i-Y_i)\beta_k .
\]
Then the positive and negative components of the gradient can be written as
\begin{align}
\left[
\nabla_{W_{ik}}\mathcal{J}_{\mathbf{W}}
\right]_{+}
&=
\sum_{j=1}^{N} H_{kj}
+
\alpha\max(0,G_{ik}), \nonumber \\
\left[
\nabla_{W_{ik}}\mathcal{J}_{\mathbf{W}}
\right]_{-}
&=
\sum_{j=1}^{N}
H_{kj}
\frac{X_{ij}}{(\mathbf{W}\mathbf{H})_{ij}}
+
\alpha\max(0,-G_{ik}).
\label{app:eq:w_positive_negative}
\end{align}
The corresponding multiplicative update is obtained by scaling the current
value of \(W_{ik}\) by the ratio of the negative and positive components:
\begin{equation}
W_{ik}
\leftarrow
W_{ik}
\frac{
\sum_{j=1}^{N}
H_{kj}
\frac{X_{ij}}{(\mathbf{W}\mathbf{H})_{ij}}
+
\alpha\max(0,-G_{ik})
}{
\sum_{j=1}^{N} H_{kj}
+
\alpha\max(0,G_{ik})
}.
\label{app:eq:w_update_no_gamma}
\end{equation}
In implementation, a stabilising scaling constant \(\gamma>0\) is added to
both the numerator and denominator:
\begin{equation}
W_{ik}
\leftarrow
W_{ik}
\frac{
\sum_{j=1}^{N}
H_{kj}
\frac{X_{ij}}{(\mathbf{W}\mathbf{H})_{ij}}
+
\alpha\max(0,-G_{ik})
+
\gamma
}{
\sum_{j=1}^{N} H_{kj}
+
\alpha\max(0,G_{ik})
+
\gamma
}.
\label{app:eq:w_update}
\end{equation}

The update in Eq.~\eqref{app:eq:w_update} preserves nonnegativity because
it rescales \(W_{ik}\) by a nonnegative ratio. Unlike the standard KL-NMF
updates, however, the supervised update for \(\mathbf{W}\) is based on
positive-negative gradient splitting for the joint objective and should not
be interpreted as a strict majorisation-minimisation update for the full
loss. It is used here as a stabilised multiplicative block update within
the alternating optimisation scheme.

\section{Additional simulation details}
\label{app:simulation_details}

This section gives the full data-generating mechanism used in the
simulation study. The simulation is designed as a controlled topic-driven
setting. It is not intended to reproduce all empirical features of online
reviews, such as sentiment asymmetry, repeated phrases, platform-specific
selection effects, or heterogeneous writing styles. Instead, it specifies a
setting in which the rating signal is known to arise from latent
document-topic proportions.

The data-generating mechanism can be summarised as
\[
\mathbf{W}_{\mathrm{true}},\mathbf{H}_{\mathrm{true}}
\longrightarrow
\mathbf{X},
\qquad
\mathbf{W}_{\mathrm{true}},\boldsymbol{\beta}_{\mathrm{true}}
\longrightarrow
\mathbf{Y}.
\]
Thus, both the observed document-term matrix and the rating outcome share
the same latent document-topic structure, while the ratings are not
generated directly from individual word counts.

The simulation proceeds in five steps. First, we generate a true topic-word
distribution matrix \(\mathbf{H}_{\mathrm{true}}\). Secondly, we generate a
document-topic proportion matrix \(\mathbf{W}_{\mathrm{true}}\). Thirdly,
we generate a document-term matrix \(\mathbf{X}\) from
\(\mathbf{W}_{\mathrm{true}}\mathbf{H}_{\mathrm{true}}\). Fourthly, we
generate a topic-rating coefficient vector
\(\boldsymbol{\beta}_{\mathrm{true}}\). Finally, we generate the rating
outcome \(\mathbf{Y}\) from \(\mathbf{W}_{\mathrm{true}}\) and
\(\boldsymbol{\beta}_{\mathrm{true}}\).

For each true topic \(k=1,\ldots,K_{\mathrm{true}}\), the topic-word
distribution is generated from
\[
\mathbf{h}_k \sim \mathrm{Dirichlet}(\alpha_w\mathbf{1}_V),
\]
where \(V\) is the vocabulary size. The parameter \(\alpha_w\) controls the
sparsity of each topic over the vocabulary. Smaller values produce sharper
topic-word distributions, while larger values produce more diffuse topics.

For each document \(d=1,\ldots,N\), the document-topic proportion vector is
generated from
\[
\mathbf{w}_d \sim
\mathrm{Dirichlet}(\alpha_{\mathrm{topic}}\mathbf{1}_{K_{\mathrm{true}}}),
\]
where \(\alpha_{\mathrm{topic}}\) controls how concentrated each document is
over the latent topics. Smaller values imply that each document is
dominated by only a few topics, while larger values produce more mixed
topic compositions.

Given \(\mathbf{H}_{\mathrm{true}}\) and \(\mathbf{W}_{\mathrm{true}}\), the
word-probability vector for document \(d\) is defined as
\[
p_{dj}^{(x)}
=
\frac{(\mathbf{w}_d^\top\mathbf{H}_{\mathrm{true}})_j}
{\sum_{\ell=1}^{V}(\mathbf{w}_d^\top\mathbf{H}_{\mathrm{true}})_\ell},
\qquad j=1,\ldots,V.
\]
The document length is generated from
\[
L_d \sim \mathrm{Poisson}(\bar{L}),
\]
and the observed word-count vector is sampled as
\[
\mathbf{x}_d \sim \mathrm{Multinomial}(L_d,\mathbf{p}^{(x)}_d).
\]
Thus, \(\mathbf{X}\) is a noisy and sparse observation of the latent topic
structure rather than a deterministic representation of it.

The rating outcome is generated directly from the latent document-topic
proportions. For each topic \(k=1,\ldots,K_{\mathrm{true}}\), we draw
\[
\beta_k \sim \mathcal{N}(0,\sigma_\beta^2),
\]
and define
\[
\eta_d=\mathbf{w}_d^\top\boldsymbol{\beta}_{\mathrm{true}},
\qquad
p_d=\mathrm{logit}^{-1}(\eta_d).
\]
The rating outcome is then sampled from
\[
Y_d \sim \mathrm{Binomial}(n,p_d).
\]
We set \(n=4\), so that \(Y_d\in\{0,1,2,3,4\}\). This corresponds to a
five-level rating scale after shifting by one unit for presentation.

The baseline simulation setting is reported in Table~\ref{app:tab:sim-settings}.
This configuration produces sparse short texts, clearly structured topics,
and a moderate-to-strong topic-level rating signal.

\begin{table}[!t]
\caption{Simulation settings for the topic-driven data-generating mechanism.\label{app:tab:sim-settings}}%
\begin{tabular*}{\textwidth}{@{\extracolsep{\fill}}llp{0.45\textwidth}@{\extracolsep{\fill}}}
\toprule
Quantity & Value & Role \\
\midrule
\(K_{\mathrm{true}}\) & 5 & True number of generating topics \\
\(N\) & 5000 & Number of documents \\
\(V\) & 2000 & Vocabulary size \\
\(\bar{L}\) & 30 & Average document length \\
\(\alpha_{\mathrm{topic}}\) & 0.05 & Document-topic sparsity \\
\(\alpha_w\) & 0.01 & Topic-word sparsity \\
\(\sigma_\beta\) & 2.0 & Strength of the topic-rating signal \\
\(n\) & 4 & Binomial size for the five-level rating scale \\
\bottomrule
\end{tabular*}
\end{table}

\section{Additional empirical data summaries}
\label{app:empirical_data_summaries}

This section provides additional descriptive summaries for the TripAdvisor
review sample used in the empirical application. These summaries are not
used as model inputs beyond the text and rating variables, but they provide
context for the rating distribution and vocabulary composition of the
corpus.

\begin{figure}[!t]
\centering
\includegraphics[width=0.70\textwidth]{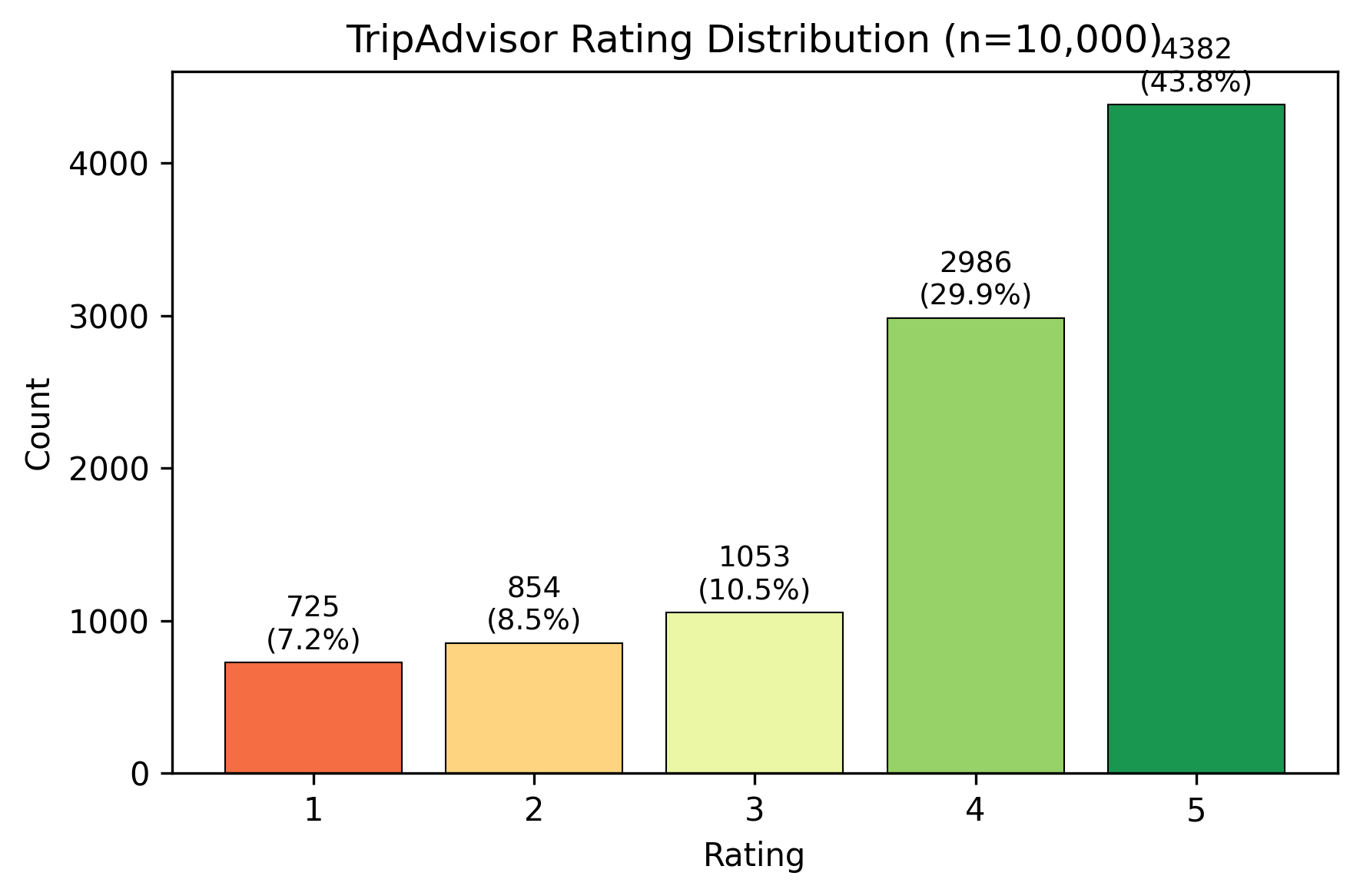}
\caption{Rating distribution of the TripAdvisor review sample, with
\(N=10{,}000\).\label{app:fig:trip-rating-dist}}%
\end{figure}

Figure~\ref{app:fig:trip-rating-dist} shows the rating distribution for
the TripAdvisor sample. The distribution is concentrated at the upper end of
the five-point scale: 4-star and 5-star reviews account for 73.7\% of the
sample. This concentration reduces the effective variation in the outcome
and makes the prediction task different from the controlled simulation
setting.

\begin{figure}[!t]
\centering
\includegraphics[width=0.70\textwidth]{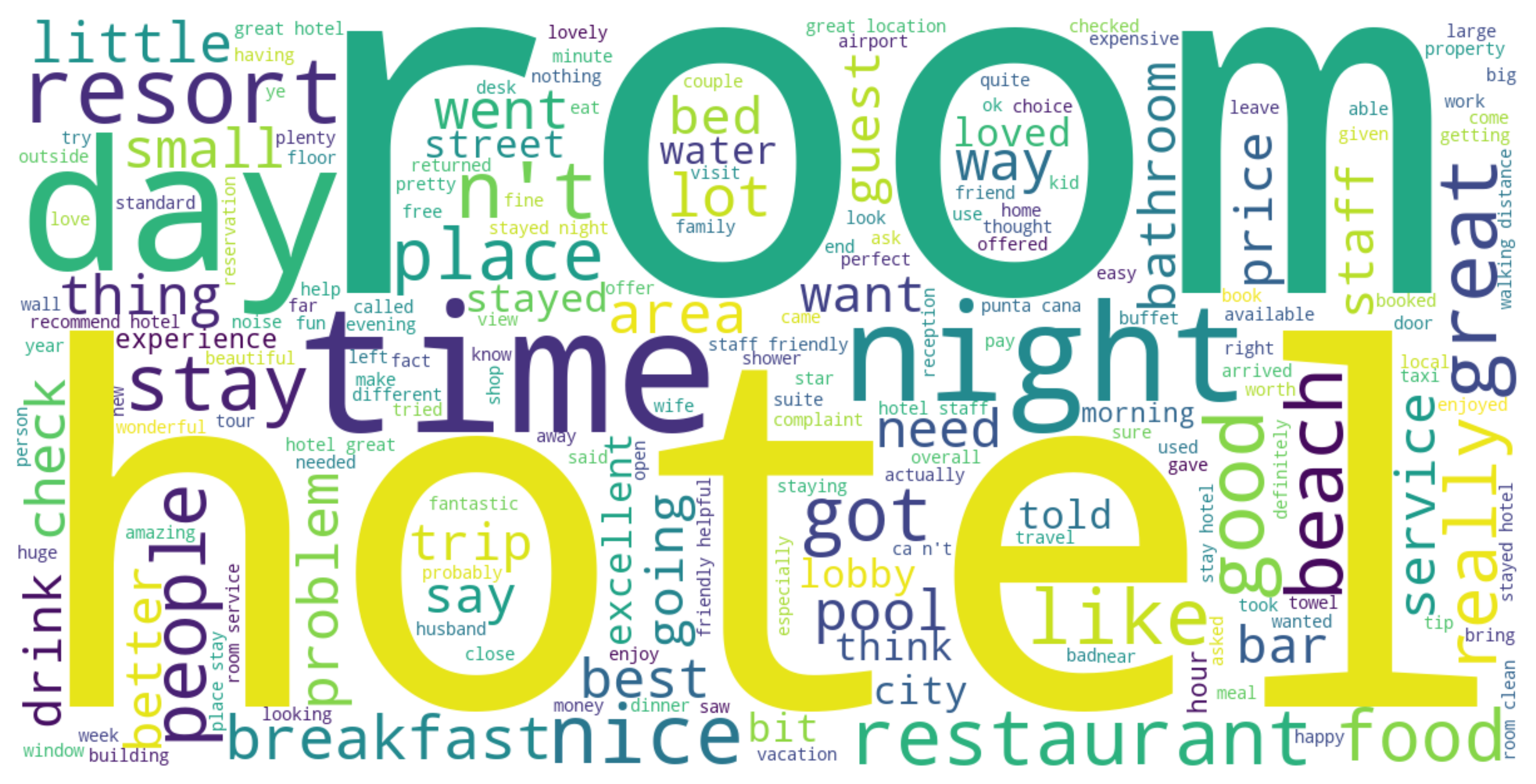}
\caption{Overall word cloud of the TripAdvisor review corpus.
\label{app:fig:trip-wordcloud}}%
\end{figure}

Figure~\ref{app:fig:trip-wordcloud} gives an overall word cloud for the
TripAdvisor review corpus. The most frequent terms are related to hotel and
service experience, including words such as room, hotel, staff, breakfast,
food, location, and stay. The word cloud is intended only as a descriptive
summary of the corpus vocabulary; the topic-level interpretation in the main
analysis is based on the fitted JBNMF topics rather than on raw word
frequency.

\section{Additional parameter-search results for the empirical application}
\label{app:empirical_parameter_search}

This section reports the full reduced search results for JBNMF on the main
TripAdvisor sample over the fitted number of topics \(K\) and the
supervision weight \(\alpha\). The numerical stabilisation parameter and
regularisation parameter are fixed at \(\gamma=400\) and \(\lambda=0.1\),
respectively.

\begin{figure*}[!t]
\centering
\begin{minipage}{0.49\textwidth}
\centering
\includegraphics[width=\linewidth]{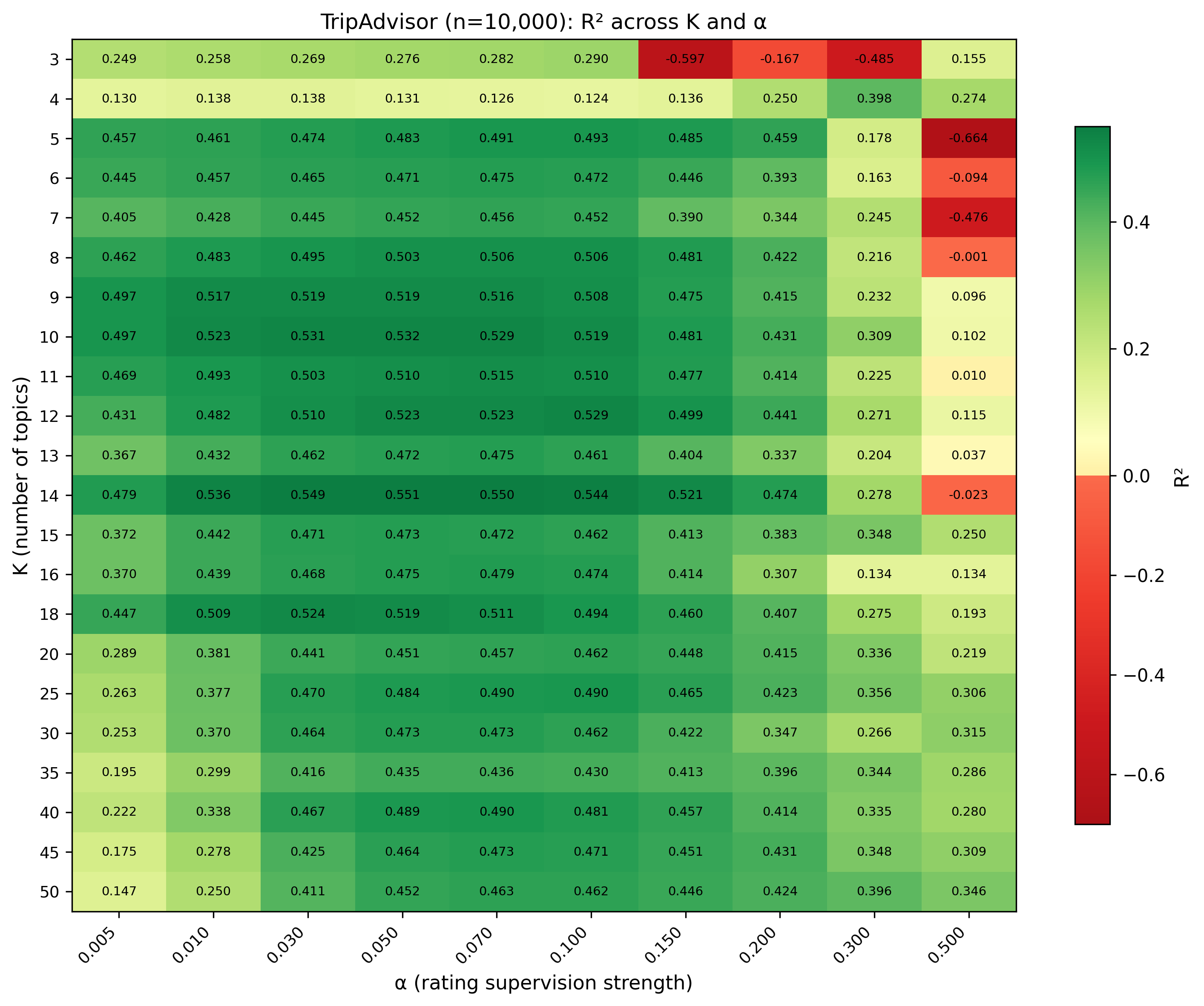}
\textbf{(a)} Test \(R^2\)
\end{minipage}
\hfill
\begin{minipage}{0.49\textwidth}
\centering
\includegraphics[width=\linewidth]{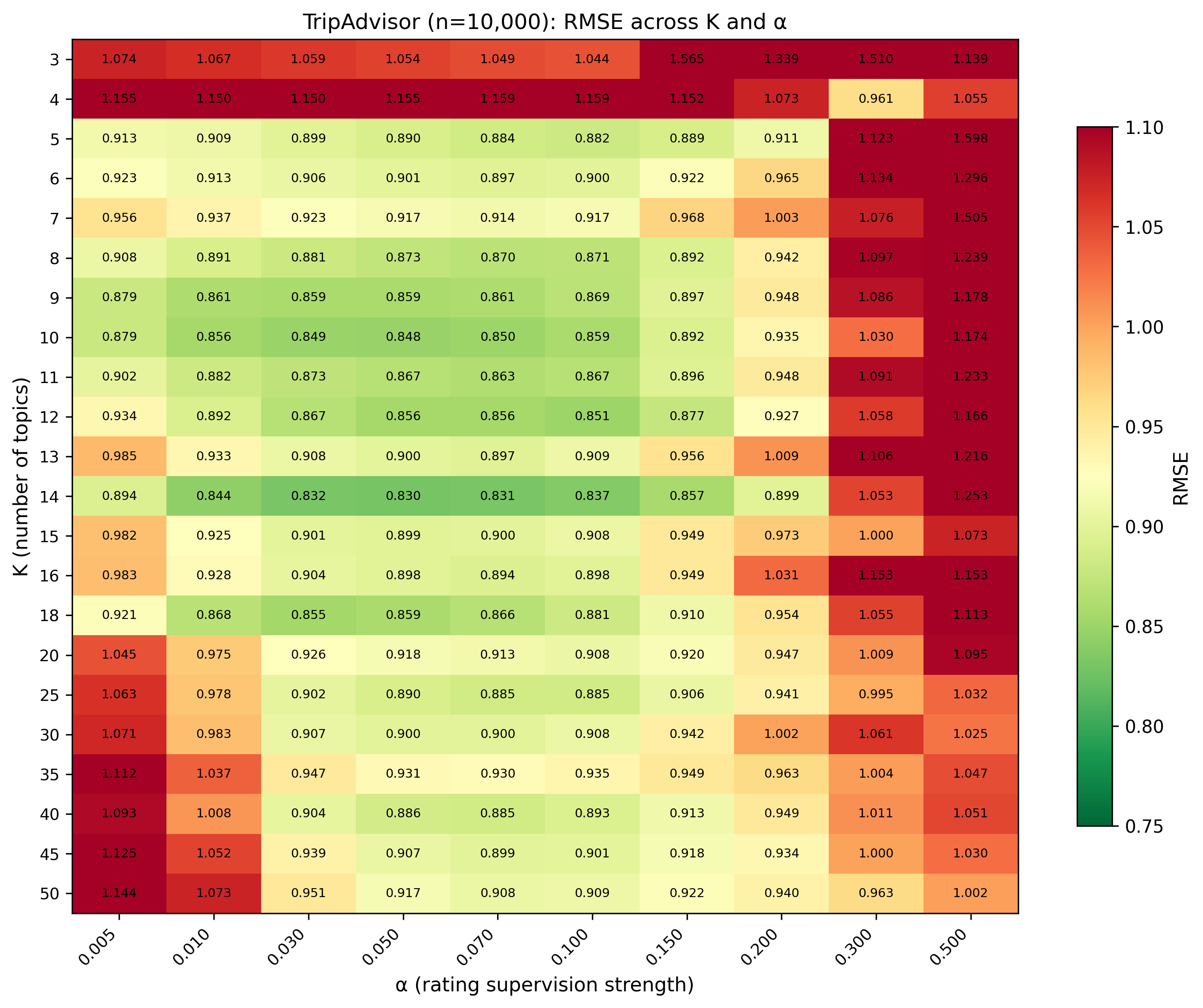}
\textbf{(b)} Test RMSE
\end{minipage}
\caption{Predictive performance of JBNMF on the TripAdvisor sample under
the reduced search over \(K\) and \(\alpha\), with \(N=10{,}000\),
\(\gamma=400\), and \(\lambda=0.1\). Panel (a) reports test \(R^2\), and
panel (b) reports test RMSE.\label{app:fig:trip-parameter-search}}%
\end{figure*}

Figure~\ref{app:fig:trip-parameter-search} shows the test \(R^2\) and RMSE
over the searched grid. The best configuration is obtained at \(K=14\) and
\(\alpha=0.05\), giving \(R^2=0.551\) and RMSE \(=0.830\). The
higher-performing region is concentrated around moderate values of \(K\) and
small to moderate values of \(\alpha\), whereas larger values of \(\alpha\)
are less stable and often lead to weaker predictive performance.

\section{Additional overfitting analysis}
\label{app:additional-overfitting}

\begin{figure*}[!t]
\centering
\includegraphics[width=\textwidth]{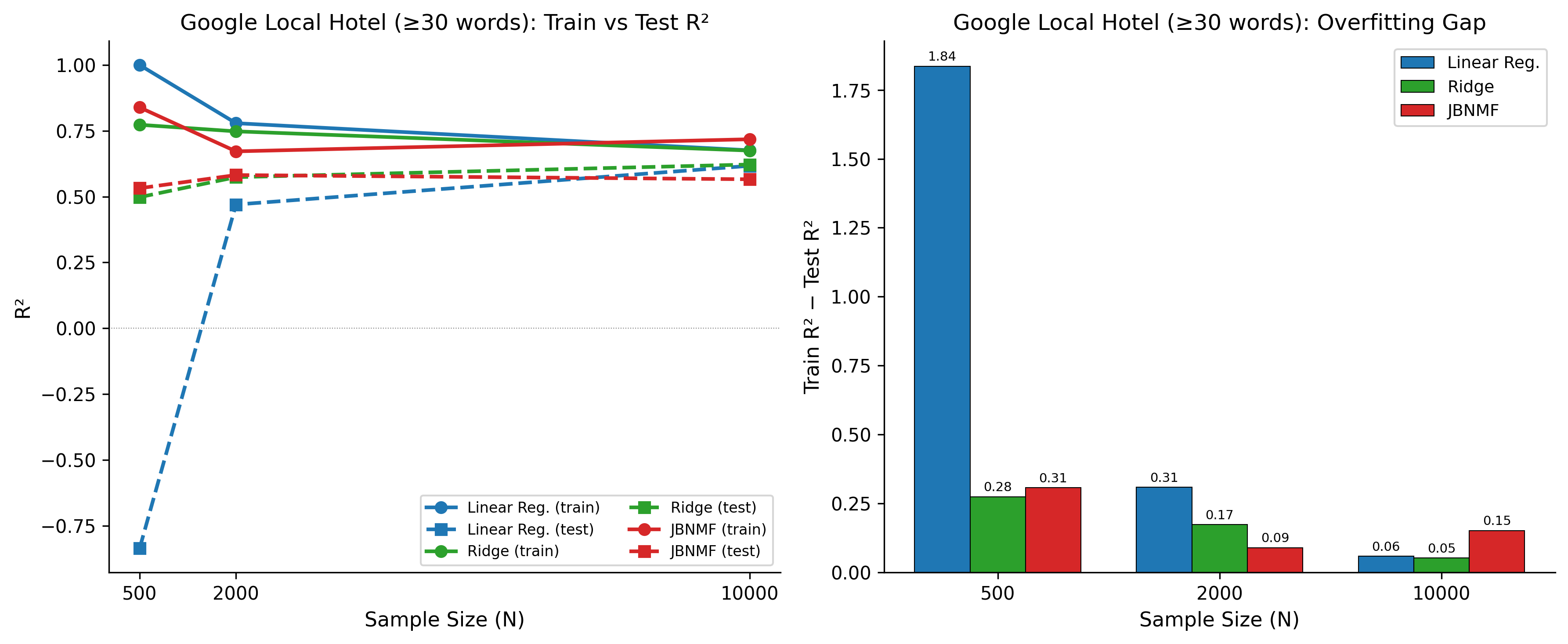}
\caption{Additional training and test \(R^2\) comparison on a Google Local
Hotel subset. The right panel reports the overfitting gap
\(\Delta R^2=R^2_{\mathrm{train}}-R^2_{\mathrm{test}}\).
\label{app:fig:hotel-overfitting}}%
\end{figure*}

Figure~\ref{app:fig:hotel-overfitting} reports the same train-test \(R^2\)
comparison on an additional Google Local Hotel subset. This subset is used
as an additional robustness check rather than as the primary empirical
setting. The overall pattern is consistent with the TripAdvisor analysis:
unregularised linear regression has the largest small-sample train-test gap,
while ridge regression and JBNMF show more stable held-out performance.

\end{document}